\RequirePackage{lineno}
\documentclass[12pt]{article}
\usepackage{graphicx,color,epstopdf,grffile,upgreek}
\usepackage[bookmarks,pdfhighlight=/O,pdfstartview=FitH]{hyperref}
\usepackage{amsfonts,mathrsfs,amssymb,amsmath,amsopn,amsthm,bm,bbm,slashed,qmech}
\title{%Review of opportunities for 
PANDA as midrapidity detector for a future HESR Collider at FAIR}

\author{Leonid Frankfurt$^1$, Mark Strikman$^2$, Alexei Larionov$^{9}$,\\ 
Andreas Lehrach$^{3,6}$, Rudolf Maier$^{3,6}$, Hendrik van Hees$^{7}$,\\ 
Christian Spieles$^7$, Volodymyr Vovchenko$^{5,7}$, Horst Stoecker$^{5,7,8}$
\\ $^1${\small \it Sackler School of Exact Sciences, Tel Aviv University, Tel Aviv, Israel}
\\ $^2${\small \it Pennsylvania State University, University Park, PA, USA}
\\ $^3${\small \it Institut f\"ur Kernphysik, Forschungszentrum J\"ulich, D-52425 J\"ulich, Germany}
\\ $^4${\small \it National Research Center ``Kurchatov Institute'', 123182 Moscow, Russia}
\\ $^5${\small \it Frankfurt Institute for Advanced Studies, Giersch Science Center,}
\\{\small \it D-60438 Frankfurt am Main, Germany}
\\ $^6${\small \it JARA-FAME (Forces and Matter Experiments), Forschungszentrum J\"ulich }
\\{\small \it and RWTH Aachen University, Germany}
\\ $^7${\small \it Institut f\"ur Theoretische Physik,
Goethe Universit\"at Frankfurt,}
\\{\small \it D-60438 Frankfurt am Main, Germany}
\\ $^8${\small \it GSI Helmholtzzentrum f\"ur Schwerionenforschung GmbH,}
\\{\small \it D-64291 Darmstadt, Germany}
\\ $^9${\small \it Institut f\"ur Theoretische Physik, Universit\"at Giessen, D-35392 Giessen
, Germany}
}

\date{\today}

\begin{document}
\maketitle
\begin{abstract}
  Exciting new scientific opportunities are presented for the PANDA
  detector at the High Energy Storage Ring in the redefined
  $\overline{\text{p}} \text{p}(A)$ collider mode, HESR-C, at the
  Facility for Antiproton and Ion Research (FAIR) in Europe.  The high
  luminosity, $L \sim 10^{31}$ cm$^{-2}$ s$^{-1}$, and a wide range of
  intermediate and high energies, $\sqrt{s_{\text{NN}}}$ up to 30 GeV
  for $\overline{\text{p}} \text{p}(A)$ collisions will allow to explore
  a wide range of exciting topics in QCD, including the study of the
  production of excited open charm and bottom states, nuclear bound
  states containing heavy (anti)quarks, the interplay of hard and soft
  physics in the dilepton production, probing short-range correlations
  in nuclei, and the exploration of the early, complete
  $\overline{\text{p}}$-p- annihilation phase, where an intially pure
  Yang-Mills gluon plasma is formed\footnote{The present article is
    based on the article {\protect \cite{Frankfurt:2018msx}} to be
    published in the Springer-Nature FIAS Interdisciplinary Science
    series.}.
\end{abstract}

\section{Introduction}

The experimental discovery of charmonium
\cite{Augustin:1974xw,Aubert:1974js} and bottomonium \cite{Herb:1977ek}
in $\e^+\e^-$ and p$A$ collisions suggests that hadrons containing heavy
quarks can be investigated in hadronic processes, where a dense, strongly
interacting medium is formed. It can be particularly useful to
study the annihilation of antiprotons on free protons and baryons bound
in nuclei in $\overline{\text{p}} \text{p}(A)$ collisions, in both collider
and fixed-target experiments at collision energies of
$\sqrt{s}=2$-$200 \, \GeV$.

A unique opportunity to do this in the near future is provided by the
Facility for Antiproton and Ion Research
(FAIR)\footnote{\href{https://fair-center.eu/}{\texttt{https://fair-center.eu/}}},
with the PANDA
detector\footnote{\href{https://panda.gsi.de/oldwww/}{\texttt{https://panda.gsi.de/oldwww/}},
  \href{https://panda.gsi.de/frontpage}{\texttt{https://panda.gsi.de/frontpage}}}
at the high-energy storage ring (HESR).  This concerns both, the
presently developed HESR fixed-target mode at $\sqrt{s}<5.5 \, \GeV$,
and a future collider mode at $\sqrt{s}< 30\, \GeV$, with PANDA as
midrapidity detector, see the Appendix for a brief description of the
detector.

\begin{figure}[t]
\centering
 \includegraphics[width=1.0 \linewidth]{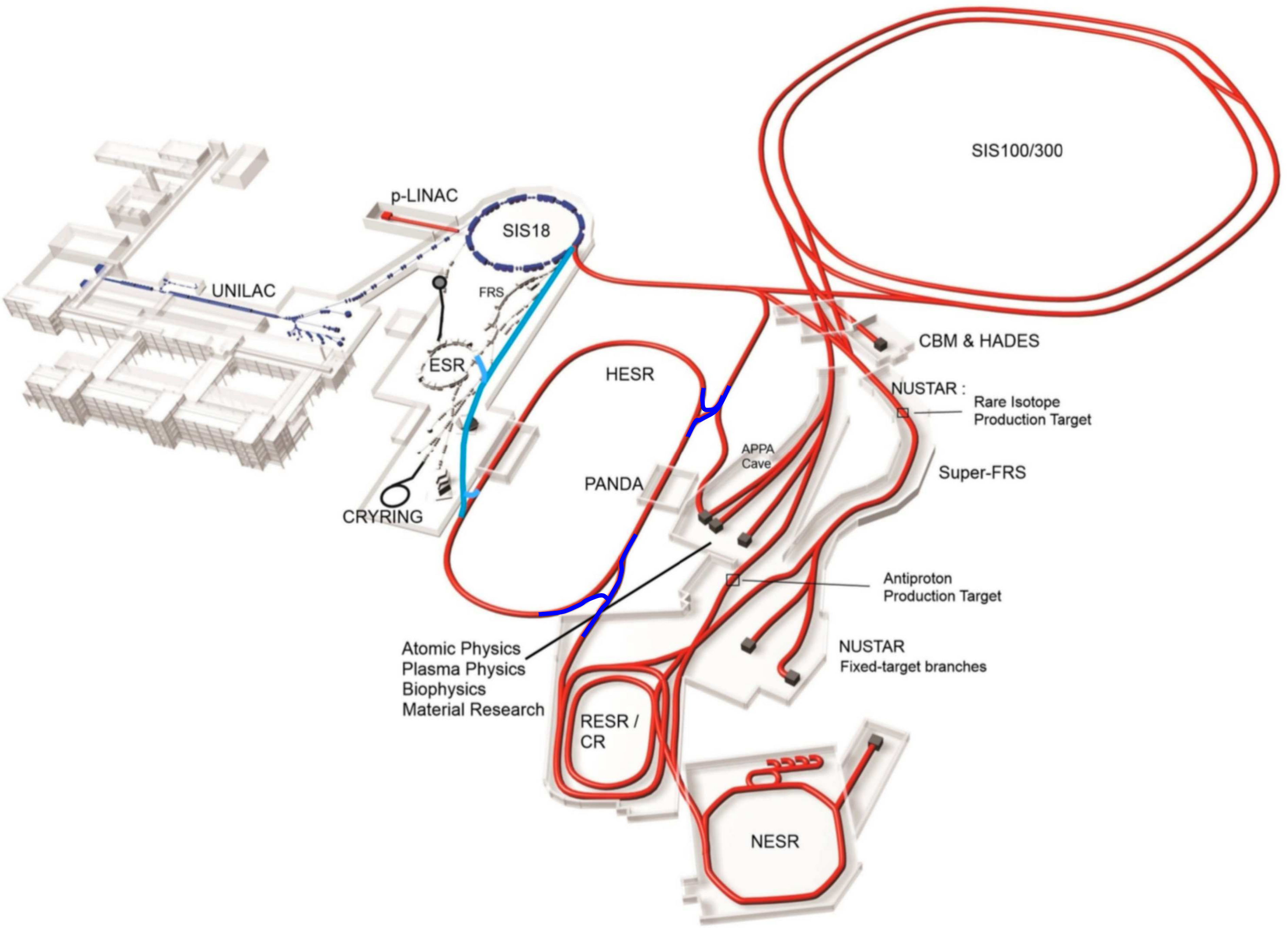}
 \caption{The existing GSI facility is shown on the left, in dark-blue
   the existing GSI accelerator facilities and ion sources. On the right
   the new FAIR complex is displayed in red with its 100 and 300 Tm
   super-conducting double-ring synchrotrons SIS100 and SIS300 (Fig.\
   adapted from \cite{Stocker:2015cva}).}
\label{fair}
\end{figure}

The collider mode would only need additional proton injection-/beam
transfer lines from the SIS18 directly into HESR, as discussed in
Ref.~\cite{Stocker:2015cva}. The proton beamline which shall shoot ions
and protons clockwise directly into the HESR-Collider shall come down
from the transfer beam line connecting the SIS18 with the SIS100. Two
switches in both directions for clockwise and counterclockwise (and vice
versa) rotating p and $\overline{\text{p}}$ beams can be realized in a
way similar to the double switch shown as the blue beamline in Fig.\
\ref{fair}, so that both the antiprotons and the protons can be made to
circulate the HESR-C counterclockwise and/or clockwise, whereas the
PANDA detector can remain nearly unchanged. This allows to eliminate the
need for an additional forward arm spectrometer.
Fig.\ \ref{fair} also shows the detector areas for
\begin{enumerate}
\item atomic and plasma physics, biomedical and material sciences (APPA), 
\item the relativistic nucleus-nucleus collision experiments CBM and HADES, 
\item the radioactive beam facility (NUSTAR),  
\item the hadron physics detector system PANDA as well as the rare- isotope and antiproton production targets with storage rings. 
\end{enumerate}
Proposed beamlines for transport of protons and ions from SIS18 directly
to the High Energy Storage Ring HESR~\cite{Stocker:2015cva} and of
antiprotons and exotic ions from HESR to the ESR/CRYRING are shown in
addition (light blue). To control the systematics and to cover the full
range of kinematic acceptance with the PANDA setup it is important that
for every injection point of protons and antiprotons into the HESR-C,
switches with two injection directions will be installed, such as shown
in the dark-blue beamline, so that both the antiprotons and the protons can
be made to circulate the HESR-C counterclockwise and/or clockwise.

After the start phase the facility will be
completed by the experimental storage rings NESR (New Experimental
Storage Ring) and RESR (Recuperated Experimental Storage Ring) enhancing
the capabilities of secondary beams and by the superconducting
synchrotron SIS300 providing true parallel operation of experimental
programs as well as particle energies twentyfold higher compared to
those achieved so far at GSI. Additionally, asymmetric HESR collider
schemes with somewhat lower center-of-mass energies have been discussed
in detail \cite{Barone:2005pu,pax:2006.asy2,Lehrbach:2007.asy3}. It may
also be feasible to study $\overline{\text{p}}A$ collisions for
$\sqrt{s_{NN}}$ of up to 19\,GeV with interesting physics opportunities
\cite{Mishustin:1993qg,Larionov:2008wy}.

Luminosities of up to $5\cdot 10^{31} \, \text{cm}^{-2}\text{s}^{-1}$
can be reached at $\sqrt{s} \simeq 30 \,\GeV$ in the symmetric
$\overline{\text{p}} \text{p}$ collider mode at the
HESR~\cite{Bradamante:2005wk,Lehrach:2005ji}.  The collision scheme of
twelve proton bunches colliding with the same amount of antiproton
bunches has to be adapted to the HESR. This modification of the HESR
requires a second proton injection, the Recuperated Experimental Storage
Ring (RESR), the 8\,GeV electron cooler and a modification of the PANDA
interaction region.

Besides the deceleration of rare-isotope beams, the RESR storage ring
also accumulates high-intensity antiprotons, via the longitudinal
momentum stacking with a stochastic cooling system \cite{Beller:2006sx}.
This is achieved by injecting and pre-cooling the produced antiprotons
at 3\,GeV in the Collector Ring (CR) storage
ring.

The antiproton beam intensities in the HESR proton-antiproton collider
version require a full-energy electron cooler (8\,MeV) to avoid beam
emittance growth, which results in a decreased luminosity during the
cycle. The Budker Institute of Nuclear Physics (BINP) presented a
feasibility study for magnetized high-energy electron cooling. An
electron beam up to 1\,A, accelerated in dedicated accelerator columns
to energies in the range of 4.5-8\,MeV has been proposed.  For the FAIR
full version, it is planned to install the high-energy electron cooler
in one of the HESR straight sections
\cite{Parkhomchuk:2004jc,Reistad:2006vm,Kamerdzhiev:2014yza}.

In the fixed-target mode, at $E_{\text{kin}}=4$-$10 \, \GeV$, it will be
possible to perform complementary measurements of the cross section of
charmonium interactions with nuclear matter with the PANDA detector.

A conservative estimate of the ${\bf \overline{\text{p}}}$ luminosities
which can be reached at the startup phase without RESR is
$4 \cdot 10^{30} \, \text{cm}^{-2} \text{s}^{-1}$. We will use it below
in our estimates assuming a one-year run ($10^7 \text{sec}$).  Energies
of $\sqrt{s}$ up to $30 \,\GeV$ could be reached and it may also be
feasible to study $\overline{p} A$ collisions for $\sqrt{s_{NN}} $ of up
to 19 GeV.  In the present work we outline how the collider at the HESR
machine will extend the scope of the PANDA project, with a focus on a
few highlights. In Sections 2 and 3 we discuss the potential of the
${\bf \text{p} \overline{\text{p}}}$ collider to provide new information
in the field of heavy quark physics, with some attention devoted to the
possible discovery of new states. Some other related opportunities are
outlined in Sec.\ 4.  Sec.\ 5 presents a number of additional physics
topics that could be explored in both $\text{p} \overline{\text{p}}$ and
$\overline{\text{p}}A$ modes. These include the production of nuclear
fragments containing $\overline{\text{c}}/\overline{\text{b}}$ and/or
c/b quarks, $\text{c} \overline{\text{c}}/\text{b} \overline{\text{b}}$
pair production, color-fluctuation effects, probing pure-glue matter,
and the production of low-mass dileptons.

Concluding remarks in Sec.\ 6 close the article.

\section{Study of heavy-quark bound states}

A number of new states containing heavy quarks have been discovered
recently. These can be interpreted as pentaquark and tetraquark states
containing $\text{c} \overline{\text{c}}$ pairs.  Some of the states are
observed in decays of mesons and baryons containing b-quarks, others in
the final states of $\e^+\e^-$ annihilation, for a review see
\cite{Karliner:2015afa}.

It is widely expected that these discoveries represent just the start of
the exploration of rich new families of states containing heavy quarks.
Understanding the dynamics responsible for the existence of these states
would help to clarify many unresolved issues in the spectroscopy of
light hadrons.

A unique feature of an intermediate-energy
$\text{p} \overline{\text{p}}$ collider is that it enables the study of the
production of $Q\overline Q$ ($Q = \text{c},\text{b}$) pairs and the
formation of various hadrons containing heavy quarks rather close to the
threshold. The $\text{b} \overline{\text{b}}$ pairs are produced mostly
by annihilation of valence quarks and antiquarks, i.e.
$q\overline q \to Q\overline Q$.  The production of
$\text{c}\overline{\text{c}}$ pairs in the antiproton fragmentation
region also corresponds to this mechanism.

The invariant masses of the produced $Q\overline Q$ pairs are much
closer to the threshold in the discussed energy range than at LHC
energies. It is natural to expect that the large probability to produce
final states with small $Q\overline Q$ invariant masses should lead to a
higher relative probability to produce pentaquark and tetraquark states
compared to the one at LHC energies. Additionally, the small transverse
momenta of the $Q\overline Q $ pairs facilitate the pick up of light
quarks as compared to $Q$ or $\overline{Q}$ fragmentation. In the
antiproton-fragmentation region another $Q\overline Q $ production
enhancement mechanism, specific for antiproton interactions, is
possible: the production of $Q\overline{Q}$ pairs with large
$x \sim 0.2$-$0.4$ in the annihilation of $q \overline{q}$, which could
merge with a spectator antiquark of the antiproton carrying $x\sim 0.2$.

Another effect which can help to observe new states in medium-energy
$\overline{\text{p}}\text{p}$ collisions is the relatively low spatial
density of the system produced at moderate energies. This should
suppress final-state interactions, which could possibly hinder the
formation of weakly bound clusters of large size. An additional
advantage is a relatively small bulk hadron production which reduces the
combinatorial background significantly as compared to the LHC.

\subsection{The heavy quark production rates} 

The need for $\text{t} \overline{\text{t}}$-production cross sections has
stimulated the development of new computational techniques for
heavy-quark production in hadron-hadron collisions (see, e.g.,
Refs.~\cite{Cacciari:2005uk,Beneke:2002ph}), in particular those which
include the effects of threshold resummation.

These calculations, which are currently being validated by comparison to
data at high collision energies, predict the following cross section for
the $\text{b} \overline{\text{b}}$ pair production in
$\overline{\text{p}} \text{p}$ collisions \cite{CacciariVogt}:
\begin{equation}
  \sigma_{\text{b}\overline{\text{b}}} (\sqrt{s} = 30 \,\GeV)= 1.8 \cdot
  10^{-2} \, \upmu\text{b}.
  \label{1}
\end{equation}
This calculation has a relative uncertainty of about 30\%, and the
predicted cross-section value is about seven times higher than the
corresponding cross section for pp scattering because of the
contribution of the valence-quark valence-antiquark annihilation present
in $\overline{\text{p}} \text{p}$.  The $\text{b}\overline{\text{b}}$ cross
section per nucleon for the $15~\text{GeV} \times 6~\text{GeV}$
kinematics is a factor of 100 smaller.  Charm production is dominated by
gluon annihilation, $\text{g} \text{g} \to Q\overline Q$. This fact implies that the
corresponding cross sections are 
%close
similar
 in pp and
$\text{p}\overline{\text{p}}$ collisions, with the exception $x_F$ close to one.
%of the
%fragmentation regions. 
The experimental data in this case are rather
consistent between pp and $\overline{\text{p}} \text{p}$ and correspond to
\begin{equation}
  \sigma_{\text{c}\overline{\text{c}}}= 30\,\upmu\text{b}.
  \label{2}
\end{equation}
For the $15~\text{GeV} \times 6~\text{GeV}$ scenario the cross section per nucleon drops by a
factor of 3.

\subsection{Rate estimates}

The cross sections in Eqs. (\ref{1}) and (\ref{2}) correspond to
significant event rates for one year ($10^7 \, \text{s}$) of running at
a luminosity of $4\cdot 10^{30}\,\text{cm}^{-2}\text{s}^{-1}$.  We find
\begin{equation}
  N_{\text{b} \overline{\text{b}}} = 10^6, \quad N_{\text{c} \overline{\text{c}}} = 10^9.
\end{equation}
These numbers can be easily rescaled for a run at a different
luminosity, if required.

%%% rewrote below - please check
In b-mesons and baryons b-quarks carry most of the hadron light cone
fraction.  In the discussed energy range b-quarks are produced with
large light cone fractions $x_{\text{b}}\sim 2m_{\text{b}}/\sqrt{s}$
comparable with those of the valence (anti)quarks, $x_q$. However the
valence quarks have a large probability to loose their light cone fraction
due to the final state interaction.  This enhances the coalescence
probability of the b-quark and the valence antiquark. One finds a broad
distribution over invariant mass of $\text{b}\bar q$, leading to the
expectation of significant formation of excited states in both the meson
and the baryon channels, as well as an enhanced probability of the
production of theexcited $\text{b} \bar{\text{b}} q \bar q$ tetraquark
or $(\text{b}\bar q)-(\bar{\text{b}} q)$ mesonic molecular states. The
probability for the formation of b$\bar q$ mesons with a given flavor is
about $30\%$ of the total cross section because of the competition
between different flavors. In the discussed processes gluon radiation is
a small correction because of the restricted phase space and the large
b-quark mass.

At the collision energies considered here, the combinatorial background is
significantly smaller than at the  LHC energies, which makes the
observation of the excited states containing heavy quarks easier.

\subsection{Hidden beauty resonance production}

In contrast to charm production, the cross section for hidden beauty
resonance production $\text{p} \overline{\text{p}} \to \chi_{\text{b}}$ may
be too low to be observed at the HESR. Within the standard quarkonium
models, where a $Q\overline Q$ pair annihilates into two gluons which
subsequently fragment into light quarks, one can estimate that this
cross section drops with $M_Q$ as
$ R_{\text{Q}}=\Gamma(\chi_{\text{Q}} \to \text{p}
\overline{\text{p}})/\Gamma_{\text{tot}}(\chi_{\text{Q}}) \propto
\alpha_s^8/M^8_Q. $ This is because this cross section is proportional
to the partial width of the decay
$\text{p} \overline{\text{p}} \to \chi_{\text{b}}$, which drops with an
increase of $M_Q$. The ratio of cross sections of beauty production
through a $\chi_{\text{b}}$ intermediate state to that for charm is
$\approx [\alpha_s(M_Q)/\alpha_s(M_c)]^8/ [M_c/M_b]^{10}$ with an
additional factor of $M_Q^{-2}$ stemming from the expression for the
resonance cross section. The suppression is due to the necessity of a
light-quark rearrangement in the wave functions of the proton and antiproton  to obtain
decent overlapping with $\chi_{\text{b}}$ states. In the
non-relativistic approximation the wave functions of $\chi_{\text{b}}$
states vanish at zero inter-quark distance. Thus the overall suppression
for the total cross section of $\chi_{\text{b}}$ production as compared
to that of $\chi_{\text{c}}$ production is approximately $ 10^{-7}$.

\section{Potential for discovery of new states}

Investigating $\text{p} \overline{\text{p}}$ collisions at moderate
energies carries specific advantages for searches of new states as the
$\text{b} \overline{\text{b}}$-production rate is relatively high, while
the overall multiplicity, which determines the background level, is
rather modest.  Also, an equal number of states containing quarks and
antiquarks is produced enabling cross checks of observations using
conjugated channels. In the following we will discuss resonances
containing heavy quarks with the understanding that everything said
equally applies to the resonances containing heavy antiquarks. Although
the rates in many cases are rather modest, we nevertheless include the
discussion of these channels in view of the possibility to have a
higher-energy collider, as discussed in the final remarks (Sec.\ 6).

\subsection{$\text{b}qq$-baryons and $\text{b}\overline q$ mesons}

The current knowledge of the spectrum of the excited states containing
b- or $\overline{\text{b}}$-quarks is very limited.  According to the
PDG \cite{Tanabashi2018}, in the $q \overline{\text{b}}$ sector there
are two states $\text{B}_{\text{J}}(5970)^+$ and
$\text{B}_{\text{J}}(5970)^0$ with unknown quantum numbers which could
be excited states of the $\text{B}^+$ and $\text{B}^0$, respectively. In
the $\text{b}qq$-sector there are two baryons $\Lambda_{\text{b}}(5912)^0$ and
$\Lambda_{\text{b}}(5920)^0$ which can be regarded as orbitally excited states of
$\Lambda_{\text{b}}^0$ and one excited $\Sigma_{\text{b}}^*$ state. This is much less in
comparison with the $\text{c}qq$ sector where five excited $\Lambda_{\text{c}}^+$
states and two excited $\Sigma_{\text{c}}$ states are listed in the Review of Particle Physics 
 (all having a   weak experimental rating  ***).
%%%Hendrik - can you please add reference to the  book 

So there are plenty of opportunities here. One noteworthy issue is the
comparison of the accuracy with which the heavy-quark limit works for
hadrons containing $\text{b}$ quarks vs. those containing $\text{c}$
quarks.

\subsection{Excited states containing $\text{b} \overline{\text{b}}$}

As argued above it is very difficult to produce bound states containing
$\text{b} \overline{\text{b}}$ in the resonance process of
$\text{p} \overline{\text{p}} $ annihilation.  Nevertheless, many of
these states, as well as other states like analogs of X, Y, Z charmonium
states, could be produced in inelastic $\text{p} \overline{\text{p}}$
interactions. This is because the invariant mass of the produced
$\text{b} \overline{\text{b}} $ system is rather close to the threshold,
and because the $\text{b} \overline{\text{b}}$ pair is produced in
association with several valence quarks and valence antiquarks which
have rather low momenta relative to the $\text{b} \overline{\text{b}}$
pair.

\subsection{Baryons and mesons containing two heavy quarks}

Since there are three valence antiquarks colliding with three valence
quarks in $\overline{\text{p}} \text{p}$, one can produce two pairs of
heavy quarks in a double quark-antiquark collision.  This entails a
possibility for producing the following baryons and mesons containing
two heavy quarks:

\textit{(i) $\mathrm{c}\mathrm{c}q$ baryons}.

At the collision energies discussed, the contribution of the
leading-twist mechanism of $2\text{g} \to Q\overline Q Q\overline Q$ for the double
heavy-quark production should be quite small as it requires very large
$x$ of the colliding partons (the situation might be less pronounced for
the case of the double $\text{c} \overline{\text{c}}$ production than for
$\text{b} \overline{\text{b}}$).  Therefore, the only effective mechanism
left is the production of two pairs of heavy quarks in two hard
parton-parton collisions (see Fig.\ref{fair}).
\begin{figure}[t]
\centering
 \includegraphics[width=.6\textwidth]{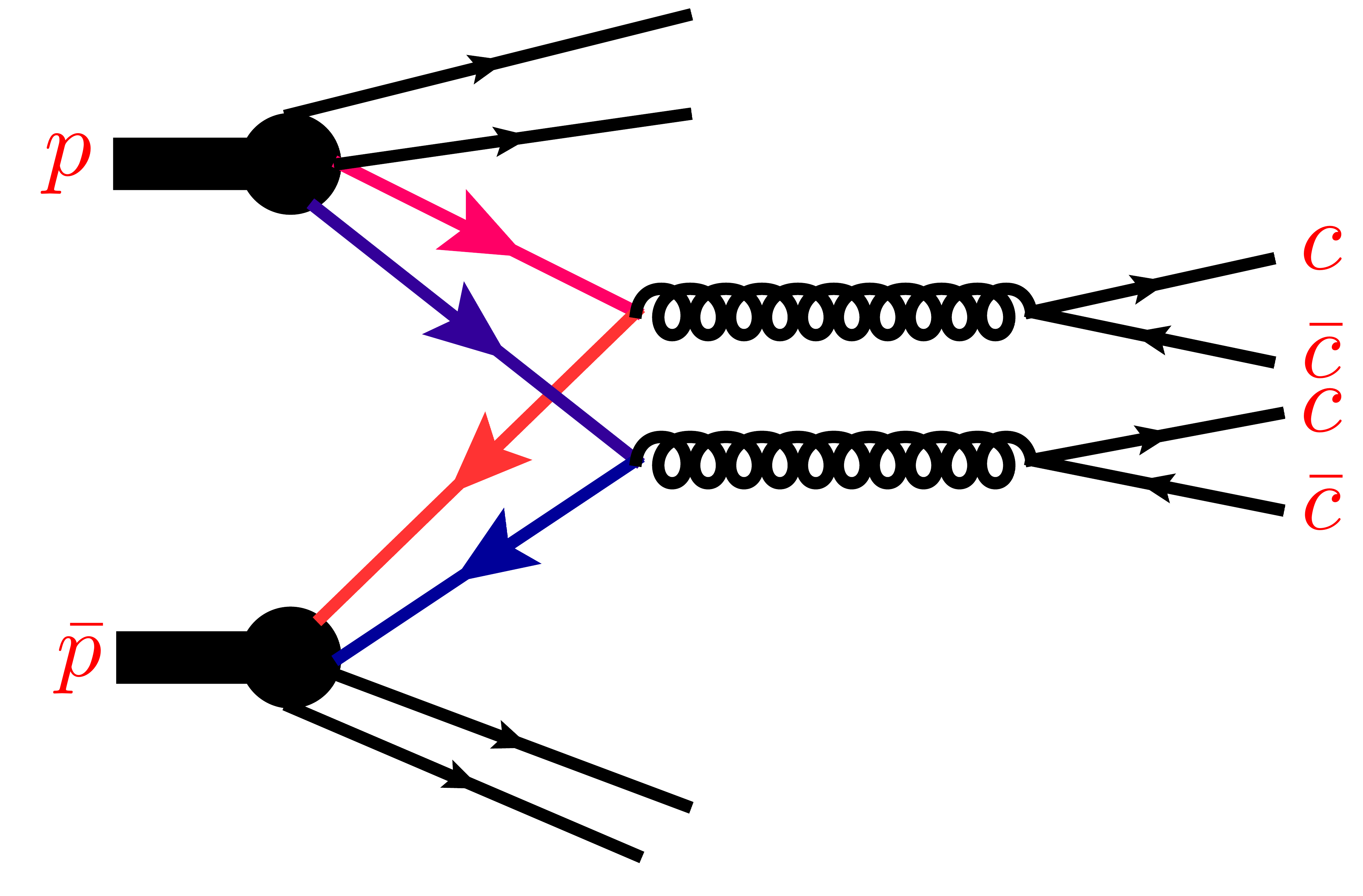}
 \caption{Double parton interaction mechanism for the production of two
   pairs of heavy quarks.}
\label{regge}
\end{figure}  

For the case of double $\text{c} \overline{\text{c}}$ pair production, one
can estimate cross sections by considering the suppression factor for
the production of the second $\text{c} \overline{\text{c}}$ pair relative to
a single $\text{c} \overline{\text{c}}$ pair.  This factor can be roughly
estimated using the high-energy experimental studies of double-parton
collisions at the Tevatron collider. One finds a probability of about
$ 10^{-3}$ for the ratio of the cross section for producing two
$\text{c} \overline{\text{c}}$ pairs relative to a single pair.
(To be conservative we took a factor of two larger value of
$\sigma_{\text{eff}} =30 \, \text{mb}$ corresponding to no correlations
of gluons in the nucleon wave function. This is a factor of two larger
than the value measured at Tevatron where small-$x$ effects reduce
$\sigma_{eff} $, for a recent review see \cite{Blok:2017alw}. )
These considerations result in
\begin{equation}
  N_{\text{c} \overline{\text{c}}, \text{c} \overline{\text{c}}} = 10^6,
\end{equation}
as an estimate for the yearly number of events with two pairs of
$\text{c} \overline{\text{c}}$. Since the available phase space is
rather modest, there is a significant probability that the relative
velocity of two c-quarks would be small, and therefore a ccq state would
be formed.

Other interesting channels are the production of an open charm-anticharm
pair plus a charmonium, and simultaneous production of two  charmoniums.

\textit{(ii) $\mathrm{bc}q$ baryons and $\mathrm{b}\overline{\mathrm{c}}$ mesons}.

The $\text{b}\overline{\text{b}} \text{c} \overline{\text{c}}$ pairs are produced
pretty close to threshold and have small relative velocities.  Hence
there is a good chance that they would form a $\text{bc}q$ baryon.
About $10^3$ events per year of running with
$\text{b} \overline{\text{b}} \text{c} \overline{\text{c}}$ could be expected
based on the double parton interaction mechanism.

\textit{(iii) $\mathrm{b}\mathrm{b}q$ baryons.}

Assuming bb is produced via double parton interaction mechanism ( Fig.\
\ref{regge}) we estimate that a production of $\sim 10^2$
$\text{b} \bar{\text{b}}\text{b} \bar{\text{b}}$ would require a
a one-year run at a much higher luminosity of
$10^{33} \,\text{cm}^{-2} \text{s}^{-1}$. The velocities of two b quarks
are expected to be close in about 1/2 of the events. So there should be
a significant chance for them to form $\text{bb}q$ baryons.
 
\textit{(iv) $\mathrm{bcc}$ baryons.}

Whether it is feasible to observe $\text{bcc}$ states requires more
detailed estimates and may depend on the structure of the three-quark
configurations in the nucleon. A naive estimate
 based on the triple parton scattering mechanism
  is that $10^2$ events
with
$\text{bcc}\overline{\text{b}} \overline{\text{c}} \overline{\text{c}}$
would be produced in a one-year run at a higher luminosity of
$10^{33} \,\text{cm}^{-2} \text{s}^{-1}$.

Note also that even though a $\text{b} \overline{\text{c}}$ meson has
been observed (although its quantum numbers are not known), there are
potentially many other states built of these quarks suggesting a rich
spectroscopy (mirroring the spectra of $\text{c} \overline{\text{c}}$-
and $\text{b} \overline{\text{b}}$-onium states).

\subsection{Summary}

To summarize, it would be possible with PANDA at the HESR-C
$\overline{\text{p}}\text{p}$ collider to discover and to study
properties of meson and baryon states containing one b quark and light
(anti) quarks, complementing the charmonium states which are planned to
be explored in the PANDA fixed-target experiment. There are also good
chances to discover the double-heavy-quark baryonic and mesonic
states. These new potential observations will allow to achieve a much
deeper understanding of the bound-state dynamics in QCD.

\section{Other opportunities}

In the last decades much effort has been devoted to studying high-energy
properties of QCD in the vacuum channel - the so-called perturbative
Pomeron. Interactions in non-vacuum channels, on the other hand, have
practically not been studied. The PANDA experiment at a collider
provides a perfect kinematic coverage to study the behavior of Regge
trajectories in both the non-perturbative regime (small Mandelstam $|t|$)
as well as the possible onset of the perturbative regime. One advantage
of the antiproton beam is the possibility to study a wide range of
baryon and meson Regge trajectories, possibly including Regge
trajectories with charmed quarks. The latter will allow to check the
non-universality of the slopes of the Regge trajectories which were
observed already at positive $t$ values.

Other possibilities include Drell-Yan-pair measurements, which at the
lower end of the discussed energy range, may be extended to the limit of
exclusive processes like
$\overline{\text{p}} \text{p} \to \upmu^+\upmu^- + \text{meson}$ which
are sensitive to generalized parton-distribution functions, etc.

Another direction of studies is the investigation of correlations
between valence (anti)quarks in (anti)nucleons using multi-parton
interactions (MPI) analogous to those shown in Fig.\ \ref{regge}. In the
discussed energy range MPI get a significant contribution from
collisions of large-$x$ partons (double Drell-Yan, Drell-Yan + charm
production, etc.), and the rate of the MPI is inversely proportional to
the square of the average distance between the valence quarks. In
particular the rates would be strongly enhanced in the case of a large
probability of (anti)quark-(anti)diquark configurations in the
(anti)nucleon.

The analysis of the production of heavy-quark pairs discussed above
would also be of great interest for the study of the multiparton
structure of nucleons.

\subsection{$\overline{\text{p}} \text{p}$ and $\overline{\text{p}} A$ elastic scattering and absorption}
\label{pbarA_el}

Recently the TOTEM collaboration at the LHC performed a high precision
measurement of the ratio of the real and imaginary of pp scattering at
$\sqrt{s}=13 $ TeV.  They found it to be significantly smaller than the
expectations based on the vacuum ("Pomeron") C-even
exchange\cite{Antchev:2017yns}. 
 Furthermore, they observed that pp and
$\text{p}\overline{\text{p}}$ elastic cross sections for
$\sqrt{s}=1.96 \,\TeV$ and  $\sqrt{s}= 2.76 \,\TeV$  respectively differ strongly in the region of the
minimum ( $-t \sim 0.7\,\GeV^2$) where the cross section is dominated by
the real part of the amplitude \cite{Antchev:2018rec}. No such significant energy dependence of the real part   is expected in the Pomeron + cuts scenario,

This seems to suggest that the contribution of the C-odd amplitude is
not as small as suggested by the existing fits based on the C-even
dominance. It was suggested to refer to the C-odd exchange as Odderon
\cite{Lukaszuk:1973nt}. In pQCD such an exchange is present as the
ladder which starts from the three gluon exchange though its absolute
rate cannot be calculated so far; pQCD model calculations lead to the
expectation that the ratio of the Odderon and Pomeron amplitudes should
drop with energy. If so, one should look for the Odderon effects at
significantly smaller energies than 2 TeV. Hence high precision
measurements of Re/Im in $\text{p}\overline{\text{p}}$ scattering at a range
of energies, and their comparison with the pp data would be highly
desirable.
 
A complementary information can be obtained from measurements in elastic
antiproton-nucleus collisions. So far such measurements were done only
at LEAR for $p_{\rm lab} < 1 \,\GeV/c$. It has turned out that owing to
the forward-peaked $\overline{\text{p}} \text{p}$ elastic scattering
amplitude the Glauber model describes LEAR data on the angular
differential cross sections of $\overline{\text{p}} A$ elastic
scattering surprisingly well. This is in contrast to the $\text{p} A$
elastic scattering where the Glauber model description starts to work
only above $p_{\rm lab} \sim 1.5$ GeV/c~\cite{Alkhazov:1978et}.

Glauber theory analysis \cite{Larionov:2016xeb} has shown that the
$\overline{\text{p}} A$ and $\text{p} A$ angular differential elastic
scattering cross sections at $p_{\rm lab} = 10 \,\GeV/c$ (fixed target
PANDA) strongly differ in the diffraction minima due to the different
ratios of the real-to-imaginary parts of $\overline{\text{p}} N$ and
$\text{p} N$ elastic scattering amplitudes. Experimental confirmation of
such a behavior would be a good validity test of the Glauber theory,
important in view of its broad applications for other reaction channels,
and of the input elementary amplitudes which are typically given by
Regge-type parameterizations. In particular, the
$\overline{\text{p}} \text{n}$ elastic amplitude is accessible only by
scattering on complex nuclei. The determination of diffractive
structures at $\overline{\text{p}}A$-collider energies would require
good transverse momentum transfer resolution $\sim 10 \, \MeV/c$ and the
capability to trigger on the events where the nucleus remains intact
(see the discussion in Sec.~\ref{new_hq_states}). Light nuclear targets
are preferred as their diffractive structures are broader in
$p_{\text{t}}$, and there is a smaller number of possible excited
states.  Spin-0 targets like $^4$He are especially suitable for these
purposes. Since $^4\text{He}$ has zero spin the depth of the minimum at
$-t\approx 0.2 \,\GeV^2$ \cite{He4} is given by the real part of the
amplitude. Note that here the contribution of the Reggeon-exchange like
$\rho$ exchange is absent in this case due to the zero isospin of the
deuteron.

A related problem is the determination of the antiproton absorption
cross section on nuclei (defined as the difference between total and
elastic cross sections). Experimental data on the antiproton absorption
cross section above LEAR energies are quite scarce, although such data
are needed for cosmic ray antiproton flux calculations
\cite{Moskalenko:2001ya}.

\subsection{Rapidity  migration  of the  fragments 
 of the projectile}

The dynamics of the fragmentation of the hadronic projectiles in the
scattering off protons and nuclei is not studied in much detail 
 especially  for  forward rapidities.
The use of antiprotons for these purposes has certain
advantages as compared to the use of protons as for large rapidity
intervals it allows to distinguish contributions of the projectile and
nucleus fragmentation. It would be possible to study how much the
fragmentation of the projectile is suppressed as a function of
multiplicity of hadrons produced at the central rapidities as well as in
the nucleus fragmentation region. Use of several nuclei would allow to
study the dependence of the rapidity distribution of
$\overline{\text{p}}$, $\overline{\Lambda}$, etc.\ as a function of the
nuclear thickness, $T(b)\int_{-\infty }^\infty \dd z \rho_A(b, z)$.  In
particular one would be able to study how the spectra of antibaryons
having one or two common antiquarks vary with the projectile.

\subsection{Coherent hypernuclei production}
\label{cohHypProd}

While ordinary $\Lambda$-hypernuclei were discovered long ago, the
$\Lambda_{\text{c}}^+$- and $\Lambda_{\text{b}}^0$-hypernuclei were
predicted in the mid-70s \cite{Tyapkin:1976,Dover:1977jw} but have not
been observed so far. However, their existence is expected based on a
number of models, e.g.\ the quark-meson coupling
model~\cite{Tsushima:2002ua}.

The processes $\overline{\text{p}} \text{p} \to \overline Y Y$, where
$Y=\Lambda, \Lambda_{\text{c}}^+$ or $\Lambda_{\text{b}}^0$, have the
lowest thresholds among all possible other channels of the respective
$\overline{\text{s}} \text{s}$, $\overline{\text{c}} \text{c}$ or
$\overline{\text{b}} \text{b}$ production channels in
$\overline{\text{p}} \text{p}$ collisions. Thus, they are preferred for
$Y$-hypernuclei production as the momentum transfer to the hyperon is
relatively small.

The coherent reactions
${}^AZ(\overline{\text{p}},\overline\Lambda)\,{}^A_\Lambda(Z-1)$ for the
different states of the hypernucleus have never being studied
experimentally. It is expected that these reactions have cross sections
of the order of a few 10\,nb at $p_{\text{lab}} \sim 20 \,\GeV/c$
\cite{Larionov:2017hcm}. Thus, they can serve as a powerful source of
$\Lambda$-hypernuclei production at the lower end of the
$\overline{\text{p}}A$-collider energies. Here, the amplitude
$\overline{\text{p}} \text{p} \to \overline{\Lambda} \Lambda$ should be
dominated by the $\text{K}^*$ (or $\text{K}^*$ Regge trajectory)
exchange.

More challenging is the coherent process
${}^AZ(\overline{\text{p}},\overline\Lambda_{\text{c}}^-)\,{}^A_{\Lambda_{\text{c}}^+}Z$
\cite{Shyam:2016uxa} where the underlying
$\overline{\text{p}} \text{p} \to \overline\Lambda_{\text{c}}^{\,-}
\Lambda_{\text{c}}^+$ amplitude is due to $\text{D}^0$ and $\text{D}^{*0}$ exchanges.
One can also think of the
${}^AZ(\overline{\text{p}},\overline\Lambda_{\text{b}}^0)\,{}^A_{\Lambda_{\text{b}}^0}(Z-1)$
coherent reaction.

\section{Unique opportunities for probing QCD properties at the $\overline{\text{p}} A$ collider}

\subsection{Space-time picture of the formation of hadrons containing heavy quarks}

The kinematics of heavy-quark-state production in collisions of
p($\overline{\text{p}}$) with proton or nuclei (neglecting Fermi motion
effects) dictates that heavy states can only be produced with momenta
\begin{equation}
p_Q >  M_Q^2/2m_Nx_q - m_Nx_q/2
\label{kin}
\end{equation}
in the rest frame of the nucleus. Here $x_q$ is the $x$ of the quark of
the nucleus involved in the production of the $Q\overline Q$ pair. For
$x_q\le 0.5$ this corresponds to a charm momentum above $4\,\GeV/c$
which is much larger than typical momenta of the heavy system embedded
in the nucleus.

However, there is a significant probability that $\D$,
$\Lambda_{\text{c}},\ldots$ hadrons slow down due to final-state
interactions. Indeed, it is expected in QCD that the interaction
strength of a fast hadron with nucleons is determined by the area in
which the color is localized. For example, the cross section of the
$\psi'$-$N$ and kaon-nucleon interaction should be comparable.
At the same time the  $\psi'$-$N$ cross section should be much larger than the  $\text{J}/\psi$-$N$   cross section,
since the color is localized in a small volume in $J/\psi$,
see for example discussion in 
\cite{Gerland:1998bz}. Also the cross sections of open charm (bottom)
interactions should be on the scale $\gtrsim 10 \,\text{mb}$.

The formation distance (coherence length) can be estimated as
\begin{equation}
l_{\text{coh}} \simeq  \gamma l_0,  
\end{equation}
where $l_0 \simeq 0.5$-$1.0 \,\fm$. For the discussed energies and the
case of scattering off heavy nuclei the condition
\begin{equation}
l_{\text{coh}} \leq R_A,  
\end{equation}
is satisfied for hadrons produced in a broad range of momenta including
the central and nucleus fragmentation region. So it would be possible to
explore the dependence of the formation time and interaction strength
on, for example, the orbital angular momentum of $\D^*$. This would
allow to test the expression for the formation distance
$l_{\text{coh}}=2p/\Delta M^2$ \cite{Dutta:2012ii}, where $p$ is the hadron
momentum and $\Delta M^2$ is typically given by the difference of masses
of the hadron and %its first excited state.
 the closest  mass states with the same quantum numbers.

Observing these phenomena and hence exploring QCD dynamics in a new
domain could be achieved by studying the $A$-dependence of charm
production at momenta $\lesssim 10 \,\GeV/c$ (in the rest frame of the
nucleus).
 
\subsection{Charm bound to nuclear fragments}%New heavy-quark states}
\label{new_hq_states}

The formation of heavy mesons or baryons, H, inside nuclei implies
final-state interactions which slow down these heavy hadrons, leading to
the production of hadrons at low momenta forbidden for scattering off a
free proton 
(cf. Eq.\ref{kin}):
\begin{equation}
p_{\text{H}} \leq (m_{\text{H}}^2-m_N^2)/2m_N.
\end{equation}
In this kinematics the slow-down may be sufficient to allow for the
production of (anti-)charm quarks embedded in nuclear fragments. The
collider kinematics would make it easier to detect decays of such nuclei
than in fixed-target setups as these nuclei would be produced with
high momenta (velocities comparable to those of ordinary nuclear
fragments). Thus the discussed HESR-C collider in the $\overline{\text{p}} A$ mode would
have a high discovery potential for observing various nuclear states
containing c and/or $\overline{\text{c}}$.

Higher luminosities and higher collider energies will allow the search for
analogous b and/or $\overline{\text{b}}$ states.

\subsection{Color fluctuations in nucleons}

At high energies hadrons are thought to be interacting with each other
in frozen configurations which have different interaction strengths --
so-called color fluctuations. One can explore these phenomena in
(anti)proton-nucleus collisions in a number of ways. Here we give as one
example the study of the interaction strength of a hadron in the case of
a configuration that contains a large-$x$ ($x \geq 0.4$) parton. One
expects that in such configurations the average interaction strength is
significantly smaller than 
$\sigma_{\text{tot}}(hN)$. 
In these configurations color
screening leads to a suppression of the gluon fields and of the
quark-antiquark sea \cite{Frankfurt:1985cv}. This picture has allowed to
explain \cite{Alvioli:2014eda,Alvioli:2017wou} strong deviations of the
centrality dependence of the leading-jet production from the geometrical
picture (Glauber model of inelastic collisions) observed at the LHC in
p-Pb collisions and at RHIC in d-Au collisions.
 
Due to a fast increase of the interaction strength for small-size
configurations with increasing energy, the strength of color
fluctuations drops at higher energies. Correspondingly color-fluctuation
effects are expected to be much enhanced at the HESR-C
$\overline{\text{p}} A$-collider energies. For example, for
$x \simeq 0.6$, the cross-section ratio, of the average strength of the
interaction of a proton in a configuration containing a quark with given x, 
$\sigma_{\text{eff}}(x)$ and $\sigma_{\text{tot}}(NN)$, is expected to be
$\sim 0.25$, while at the LHC it is $\sim 0.6$.

To observe this effect one would need to study Drell-Yan production at
large $x$. A strong reduction of hadron production in the nucleus
fragmentation region would be a strong signal for the discussed
effect. For its detailed study measurements with different nuclei would
be desirable.

\subsection{Study of short range correlations at a hadron collider}

\subsubsection{Brief summary of the studies of the short -range correlations} 

The presence of the high momentum correlated component in the nuclear
wave function was expected within microscopic theories of nuclei since
the 1960s, see for example \cite{Bethe:1956zz,
  Goldstone:1957zz,Brueckner:1958zz}.

A distinctive feature of these correlations is that the singular nature
of the short range NN interaction generates a universal (up to
normalization and isospin effects) high-momentum component in the
nucleus wave function with the fast nucleon momentum balanced by a few
nucleons, predominantly by one -- $2N$ short range correlation (SRC). In
the SRC the distances between nucleons are much smaller than on average,
leading to local densities comparable to densities in the cores of
neutron stars.

Thus the investigation of the inner structure of SRCs and the pattern of
balancing the large momentum recoil are an important step in probing the
dynamics of the cores of neutron stars in the laboratory.

A number of attempts to observe experimentally such correlations have
been undertaken and it has been repeatedly found that multistep
processes mask the contribution of the SRCs leading to a widely held
belief that SRCs cannot be observed. However, the multistep processes in
similar reactions may probe the cumulative phenomenon of sub-threshold
particle production~\cite{Gorenstein:1976zg}. This has been confirmed by
microscopic transport
calculations~\cite{Motornenko:2016sfg,Panova:2019exs} and is still to be
verified experimentally.

It has been concluded in \cite{Frankfurt:1977np,Frankfurt:1981mk} that
these failures are explained by the lack of a significant difference
between the scales of the energy-momentum of the probed bound state
and the energy-momentum transfer in the measured process.

The condition for resolving said $2N$ SRCs is that the energy- and momentum
transfer, $q$, satisfies the condition
\begin{equation}
q_0 \gg k^2/m_N,  \quad q_3 \gg 2 k,
\label{cond}
\end{equation}
where $k$ is the momentum of struck nucleon.

The first high-energy signal for the presence of SRCs in nuclei comes
from the studies of the processes
\begin{equation}
a + A \to \text{fast backward nucleon + X}, a=p,\pi, \gamma,\nu,
\label{incl}
\end{equation} 
in the kinematics where the Feynman $x_{\text{F}}$ for the nucleon scaled to A
exceeds $1 +k_{\text{F}}/m_N$, where $k_{\text{F}}$ is the nucleon Fermi
momentum.

We will discuss the case of the fragmenting nucleus viewed as a
``projectile''. In the frame where the nucleus is fast, the Feynman
variable can be thus defined as
\begin{equation}
x_{\text{F}}=\frac{p_N^z}{P_A^z},
\end{equation}
where $p_N^z$ and $P_A^z$ are the longitudinal momenta of the emitted
nucleon and the nucleus, respectively. In the case of a hadron-nucleus
interaction it is convenient to introduce a scaled variable
\begin{equation}
\alpha_N= x_{\text{F}} A. 
\end{equation}
In the reference frame, where the nucleus has large a momentum $P$,
\begin{equation}
\alpha_N= Ap_N/P.
\end{equation}
Since the produced nucleon cannot carry the momentum fraction larger
than 1 of the involved nucleon subsystem, the production of a nucleon
with $\alpha \geq j$ (where $j$ is an integer $\ge 1$) requires a process involving at least $j + 1$
nucleons.

The spectator mechanism of the production of a fast backward nucleon has
been proposed in which the projectile hadron (photon, neutrino)
interacts inelastically with one nucleon of the $2N$ SRC, carrying a
light-cone fraction $\alpha_1$ and releasing the second nucleon of the
SRC \cite{Frankfurt:1977np} with $\alpha_2 \approx 2 - \alpha_1$. (Note
here that in typical high energy interactions with a nucleon and
momentum transfer to the nucleon satisfy Eq. \ref{cond}.) The model
predicts that the spectra of nucleons from deuteron and $^4$He would be
close between $\alpha \ge 1.3$ and $\alpha \sim 1.6$ with an upper limit
determined by the contribution from $3N$ SRCs and that for $A\ge 4$ the
$\alpha$-distributions are practically the same at least up to
$\alpha \sim 2$. These expectations have been confirmed by the data; for
a summary see \cite{Frankfurt:1981mk}.

Furthermore for light nuclei the relative probability of $2N$
correlations in the nuclei and the deuteron has been determined which
has been confirmed in large-$Q^2$, $x\ge 1.4$ inclusive $A(\e,\e')$
reactions \cite{Frankfurt:1993sp} which have tested both the
universality of $2N$ correlations and the dominance of the light-cone
dynamics in the (e,e') reactions.

The following studies have focused on quasi exclusive reactions starting with
the large momentum transfer ($-t \sim 5 \,\GeV^2$) reaction
\begin{equation}
\text{p} + \text{C} \to \text{pp} + \text{backward neutron} + (A-2)^*.
\end{equation}
It has been observed that for momenta $\ge 250\,\MeV$ of the struck
nucleon in about 90\% of the scattering events of a fast proton a
backward neutron with balancing momentum is produced
\cite{Piasetzky:2006ai}. Such a dominance of neutron emission is the
natural consequence of the dominance of $I=0$, $S=1$ $NN$ SRCs in the
nucleus wave function.

Further studies at Jlab of the reactions
\begin{alignat}{2}
&\e + A \to \e + \text{forward  p} + \text{spectator n(p)} + (A-2)^*, \\
&e + A \to \e + \text{forward  n} + \text{spectator p } + (A-2)^*, 
\end{alignat}
at $x\ge 1.2$, $Q^2 \ge 1.5 \, \GeV^2$ have confirmed the dominance of
the scattering off $2N$ SRCs and tested the dominance of the pn SRCs
(see \cite{Duer:2018sxh} and references therein). Note that measurements
with proton and electron projectiles are complementary as in the case of
the proton projectile the primary interaction occurs with the forward
nucleon, and the spectator is emitted in the backward direction, while
in the electron case the kinematics has been studied in which the
electron is interacting with a backward nucleon ($\alpha > 1$) and the
spectator is emitted forward. Also, due to a stronger absorption in the
case of proton projectiles, the proton induced reactions probe lower
densities than electron-induced processes. 
The collider kinematics has a number of advantages as compared to the fixed
target setup. In particular, in the Collider setup it is rather easy to
detect all products of interactions with a SRC - as they move along the
projectile direction. On the contrary, in case of the proton projectile
the SRC constituents fly in different directions - both forward and
backward.

\textit{Comment.} The concept of the relativistic nucleon momenta within
a SRC is beyond the framework of the non-relativistic theory. Popular
model assumptions like the ones used in
\cite{Arnold:1979cg,DeForest:1983ahx} violate probability conservation
as well as the global symmetries. They also lead to the presence of
certain configurations in nuclei which cannot be knocked out from the
nucleus - an obvious contradiction with quantum mechanics. In the
relativistic limit one has to take into account the production of
virtual $N\overline N$ pairs. These effects can be adequately taken into
account in light-cone quantum mechanics of nuclei which is in a direct
correspondence with the space-time evolution of high-energy processes
(for an extensive discussion see \cite{Frankfurt:1981mk}). Note also
that high-energy processes allow to investigate modifications of quark
and gluon distributions in bound nucleons and the onset of the
transition to collective behavior of quarks and gluons which is
manifested in the EMC effect.

\subsubsection{Opportunities at Colliders}

In the collider kinematics it would be possible to study in greater
detail the dynamics of interactions of antiprotons with $2N$
correlations in the inelastic channels including shadowing due to
simultaneous interaction with two nucleons of the correlation, studying
transverse momentum balance between an $\alpha_N<1$ nucleon (fast
backward in the nucleus rest frame) and $\alpha_N>1$) nucleons
(forward-emitted in the nucleus rest frame). The selection of events
with fast backward nucleons may give unique information about the
interplay between Glauber like mechanisms of knock out of nucleons and
mechanisms involving secondary cascade interactions.

It would also be possible to study quasi-elastic channels for a range of
momentum transfers above $1 \,\GeV^2$ for the scattering off both
$\alpha_N > 1$ and $\alpha_N<1$ nucleons (in the case of medium energies only
studies of scattering off $\alpha_N<1$  moving nucleons is feasible due to the
strong suppression of the cross section for scattering at large angles).

Using hadronic projectiles may play a very important role in the studies
of higher order (in particular $3N$) SRCs. Such correlations can emerge
from an iteration of two hard $NN$ interactions or from the genuine
three-nucleon interaction. The ratio of the probability of $3N$ and $2N$
SRCs is expected to grow with nuclear density for large nucleon
momenta. Hence they could play a significant role in the cores of the
neutron stars.

The current evidence for the $3N$ SRCs comes from the studies of the
reaction (\ref{incl}) for $\alpha \ge 1.6$ where contribution of $2N$ SRCs
is suppressed, from modeling $3N$ SRCs as an iteration of the hard
interactions of two nucleons \cite{Frankfurt:1981mk}, studies of the
$^3$He wave function \cite{Sargsian:2005ru}, as well as studies of the
(e,e') reaction at $x \ge 2.5$ \cite{Frankfurt:2018gyr}.

The studied reactions do not explicitly probe $3N$ correlations as they
do not measure the final state ((e,e') reactions) or just measure one
nucleon (reaction \ref{incl}).  Obviously more exclusive measurements
are necessary. The production dynamics of a nucleon spectator from
$j >2$ SRCs is more complicated than in the case of $2N$ SRCs. In the
$2N$ case it is sufficient to transfer a large energy and momentum to
the second nucleon (\ref{cond}). However in the case of $3N$ SRCs this
is not sufficient, as the second nucleon of the correlation does not
receive a large kick, and the bond between this nucleon and the
spectator is not destroyed, leading to a significant final-state
interaction and a suppression of the yield of the fast nucleon. A much
more effective mechanism is the interaction of the projectile with two
nucleons balancing a large $\alpha $ nucleon \cite{Frankfurt:1981mk}. In
the case of hadronic projectiles such double interactions occur without
suppression since the three nucleons are close to each other. However in
DIS such interactions are strongly suppressed (if $x$ is not
small). Hence a comparison of the $\text{p} (\overline{\text{p}}) A$ and
$\gamma^*A $ production of fast backward nucleons would allow to
separate different SRCs. It would be also possible to use the
final-state information to select events in which one or two nucleons of
the target are involved in the high-energy interactions with the
projectile. In the case of quasielastic scattering at large $|t|$ one can
also separate events originating from $2N$ and $3N$ SRCs by studying how
many nucleons are balancing the nucleon with large $\alpha $ and
$p_t ^2\ge 2 \,\GeV^2$. In the case of two nucleons it should be
predominantly a pn system, for three nucleons -- ppn or nnp:
$h + A \to h(p_t \ge 1.5 \, \GeV) + \text{pn} + (A-2)^*$,
$h + A \to h(p_t \ge 1.5 GeV) + \text{ppn (nnp)}+ (A-2)^*$.

At the collider one can also study effects of nonnucleonic degrees of
freedom like $\Delta$ isobars, $N^*$, etc.

Obviously these would be challenging measurements, with their
interpretations leading to a better understanding of the dynamics of
hadron-nucleus scattering. Having complementary programs at fixed-target
and electron-ion colliders would be critical for the success of such
studies as comparisons of these seemingly different reactions would test
the factorization of observed cross sections into a product of the
elementary cross section and nuclear decay function and address the
question of the role of the f.s.i. which are very different in the
electron and hadron reactions.

\subsection{Probing pure glue matter}

One of the central questions in high-energy hadronic and nuclear
collisions is how the initially non-equilibrium system evolves towards a
state of apparent (partial) thermodynamic equilibrium at later stages of
nuclear collisions. Presently, the community favors a paradigm of an
extremely rapid (equilibration time $t_{\rm eq}$ less than $0.3\,\fm/c$)
thermalization and chemical saturation of soft gluons and light quarks.

The large gluon-gluon cross sections lead to the idea
\cite{VanHove:1974wa} that the gluonic components of colliding nucleons
interact more strongly than the quark-antiquark ones. A two-step
equilibration scenario of the quark-gluon plasma (QGP) has been proposed
in \cite{Raha:1990dn,Shuryak:1992wc,Alam:1994sc}. It has been assumed that
the gluon thermalization takes place at the proper time
$\tau_{\text{g}}<1~\textrm{fm}/c$ and the (anti)quark equilibration
occurs at $\tau_{\rm th}>\tau_{\text{g}}$. The estimates of
Refs.~\cite{Biro:1993qt,Elliott:1999uz,Xu:2004mz} show that the quark
equilibration time $\tau_{\rm th}$ can be of the order of $5\,\fm/c$.
More recent studies~\cite{Kurkela:2018oqw,Kurkela:2018xxd}, based on QCD
effective kinetic theory~\cite{Arnold:2002zm}, suggest smaller values of
$\tau_{\text{th}} \sim 1$-$2\,\fm/c$ and that a longitudinally expanding
system is not yet thermalized at chemical equilibration but exhibits a
considerable momentum anisotropy. The hydrodynamic behavior, however,
sets in before chemical equilibration, making a viscous fluid dynamical
description applicable already at the early
stages~\cite{Kurkela:2015qoa,Heller:2016rtz}. In the present
exploratory study we shall focus on the effects of chemical
non-equilibrium while preserving the ideal hydrodynamic description for
simplicity. A more quantitative future study shall take into account the
viscous corrections and effects of momentum anisotropy.

\begin{figure}[t]
\centering
\includegraphics[width=0.70\textwidth]{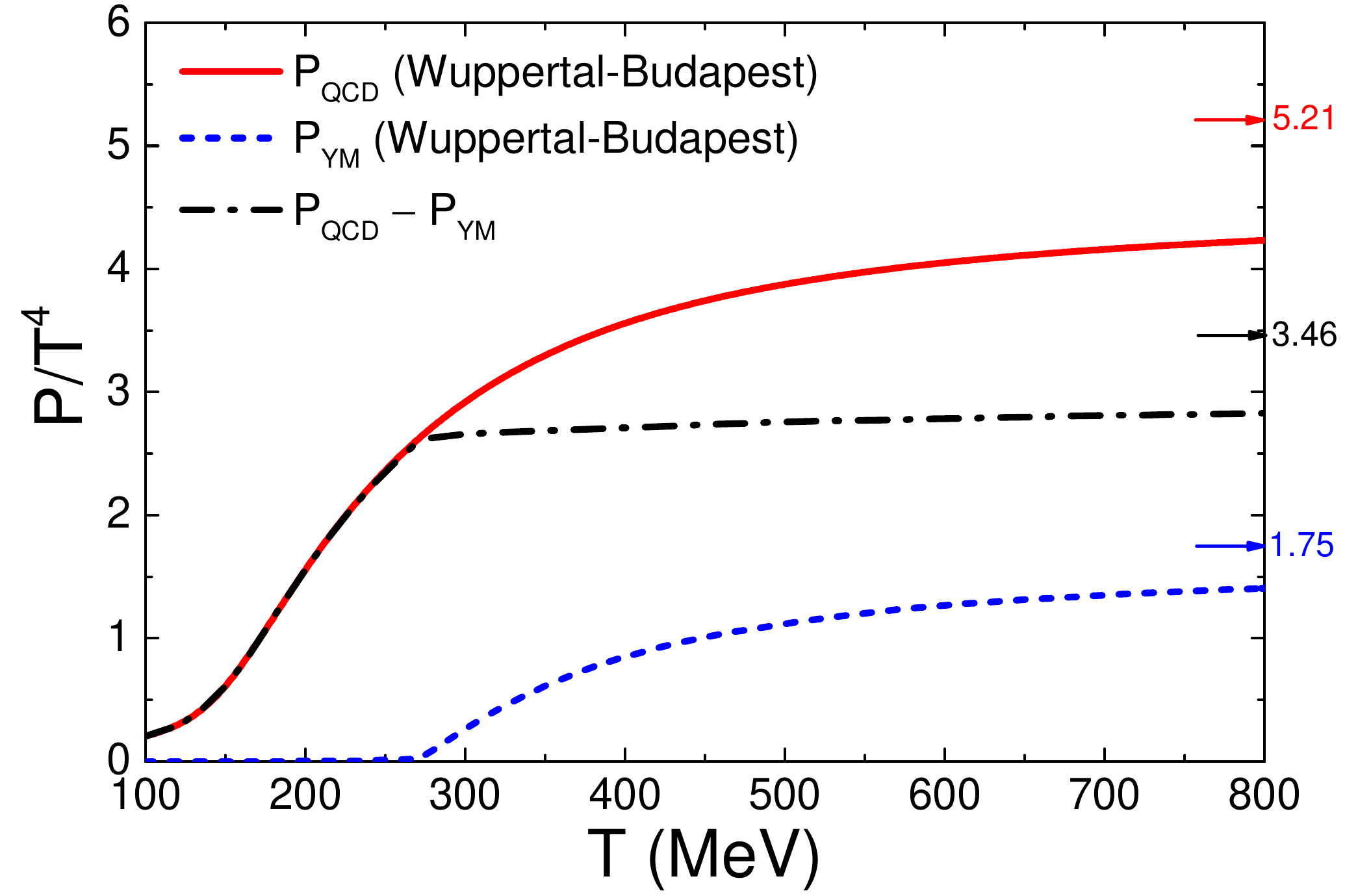}
\caption{Temperature dependence of the scaled pressure, $p/T^4$,
  obtained in lattice QCD calculations of the Wuppertal-Budapest
  collaboration for (2+1)-flavor QCD \cite{Borsanyi:2013bia} (red line)
  and for Yang-Mills matter \cite{Borsanyi:2012ve} (blue line). The
  black dash-dotted line depicts the difference between the pressure in
  full QCD and in Yang-Mills theory.}\label{fig:latticeEoS}
\end{figure}

The \emph{pure glue} scenario has been proposed for the initial state at
midrapidity in Pb+Pb collisions at the Relativistic Heavy Ion Collider
(RHIC) and Large Hadron Collider (LHC)
energies~\cite{Stoecker:2015zea,Stocker:2015nka}. According to
lattice-QCD calculations~\cite{Borsanyi:2012ve}, quarkless purely
gluonic matter should undergo a first-order phase transition at a
critical temperature $T_{\text{c}}= 270 \,\MeV$. At this temperature the
deconfined pure-glue matter transforms into the confined state of pure
Yang-Mills theory, namely into a glueball fluid. This is in stark
contrast to full QCD equilibrium with (2+1) flavors, where a smooth
crossover transition takes place (see Fig.~\ref{fig:latticeEoS} for a
comparison of the corresponding equations of state).

At $\sqrt{s_{NN}} \simeq 30\,\GeV$ $\overline{\text{p}}\text{p}(A)$
collisions can create only small systems. Baryon free matter can be
expected if the $\overline{\text{p}}\text{p}$ annihilation occurs
briefly in the initial stage of the collision. An enhanced annihilation
probability can be expected in $\overline{\text{p}}A$ collisions over
the $\overline{\text{p}}\text{p}$ collisions. However, the matter
created in $\overline{\text{p}}A$ collisions will not be net-baryon
free. In the following we focus on the particular class of
$\overline{\text{p}}\text{p}$ events where baryon free matter is
created.  Note that the described $\overline{\text{p}}\text{p}$
collisions at PANDA are quite different from the heavy-ion program of
the future compressed baryonic matter (CBM) experiment at FAIR. The
heavy-ion collisions at CBM will create excited baryon-rich QCD matter
rather than baryon-free pure glue matter in PANDA. Therefore, the PANDA
program will be complementary to the CBM experiment.

If indeed a hot thermalized gluon fluid, initially containing no
(anti)quarks, is created in the early stage of a
$\overline{\text{p}}\text{p}$ (or $\overline{\text{p}}A$ collision) at
mid rapidity, it will quickly cool and expand until it reaches a
mixed-phase region at $T = T_{\text{c}}^{\text{YM}}$. After the initial
pure gluon plasma has completely transformed into the glueball fluid,
the system will cool down further. The heavy glueballs produced during
the Yang-Mills hadronization process, where the pure glue plasma forms a
glueball fluid, will later evolve into lighter states, possibly via a
chain of two-body decays~\cite{Beitel:2016ghw}, and finally decay into
hadronic resonances and light hadrons, which may or may not show
features of chemical equilibration.

Of course, a more realistic scenario must take into account that some
quarks will be produced already before and during the Yang-Mills driven
first-order phase transition. This scenario can be modeled by
introducing a time-dependent effective number of (anti)quark degrees
of freedom, given by the time-dependent absolute quark fugacity
$\lambda_q$~\cite{Vovchenko:2015yia}:
\begin{equation}
\label{eq:fug}
\lambda_q (\tau) = 1 - \exp\left(\frac{\tau_0-\tau}{{\tau_*}}\right)\,.
\end{equation}
Here $\tau_*$ characterizes the quark chemical equilibration time, and
$\tau$ is the longitudinal proper time.

To illustrate the above considerations, we apply the (2+1)-dimensional
relativistic hydrodynamics framework with a time-dependent equation of
state, developed in Refs.~\cite{Vovchenko:2016ijt,Vovchenko:2016mtf} and
implemented it in the \texttt{vHLLE} package~\cite{Karpenko:2013wva}, to
$\text{p}\overline{\text{p}}$ collisions at HESR. The equation of state
interpolates linearly between the lattice equations of state for the
purely gluonic Yang-Mills (YM) theory~\cite{Borsanyi:2012ve}
$P_{\rm YM}(T)$ at $\lambda_q = 0$ and for the full QCD with (2+1) quark
flavors \cite{Borsanyi:2013bia} $P_{\rm QCD}(T)$ at $\lambda_q = 1$:
\begin{equation}
\begin{split}
\label{eq:interpolated}
P(T, \lambda_q) & = \lambda_q \, P_{\rm QCD} (T) + (1 - \lambda_q) \, P_{\rm YM} (T) \nonumber \\
& = P_{\rm YM} (T) + \lambda_q \, [ P_{\rm QCD} (T) - P_{\rm YM} (T) ].
\end{split}
\end{equation}
$P_{\rm YM}(T)$ and $P_{\rm QCD}(T)$ are shown in
Fig.~\ref{fig:latticeEoS}.

The hydrodynamic simulations of $\text{p} \overline{\text{p}}$
collisions at $\sqrt{s} = 30\,\GeV$ discussed below assume a hard-sphere
initial energy density profile with radius $R = 0.6 \, \fm$. The
normalization of the energy is fixed in order to yield an initial
temperature of 273\,MeV in the central cell, which is slightly above the
critical temperature of 270\,MeV. This choice is motivated by
Bjorken-model based estimates at $\sqrt{s} = 30 \,\GeV$ for small
systems~\cite{VovchenkoThesis}.

\begin{figure}[!h]
\centering
\includegraphics[width=0.49\textwidth]{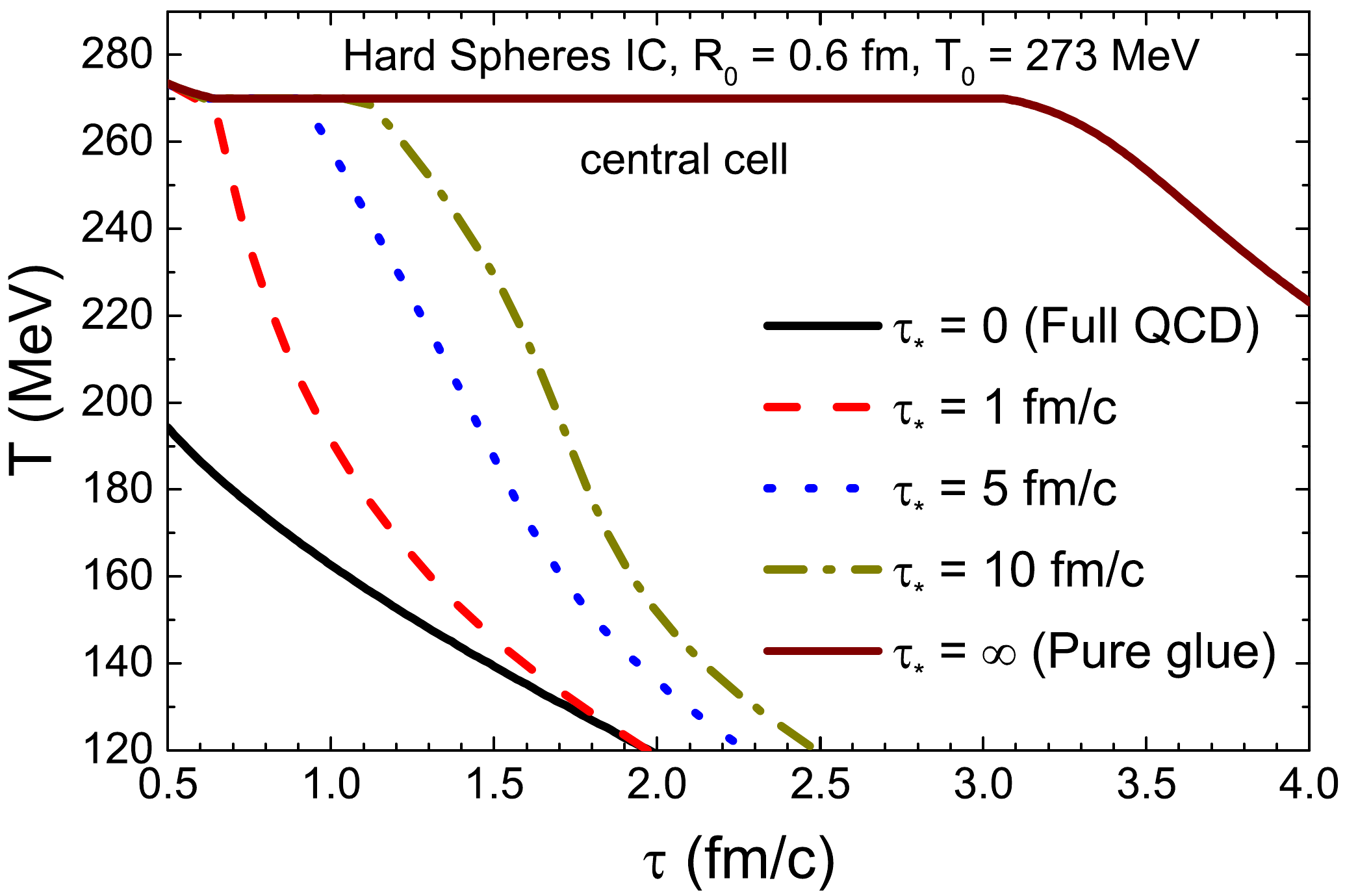}
\includegraphics[width=0.49\textwidth]{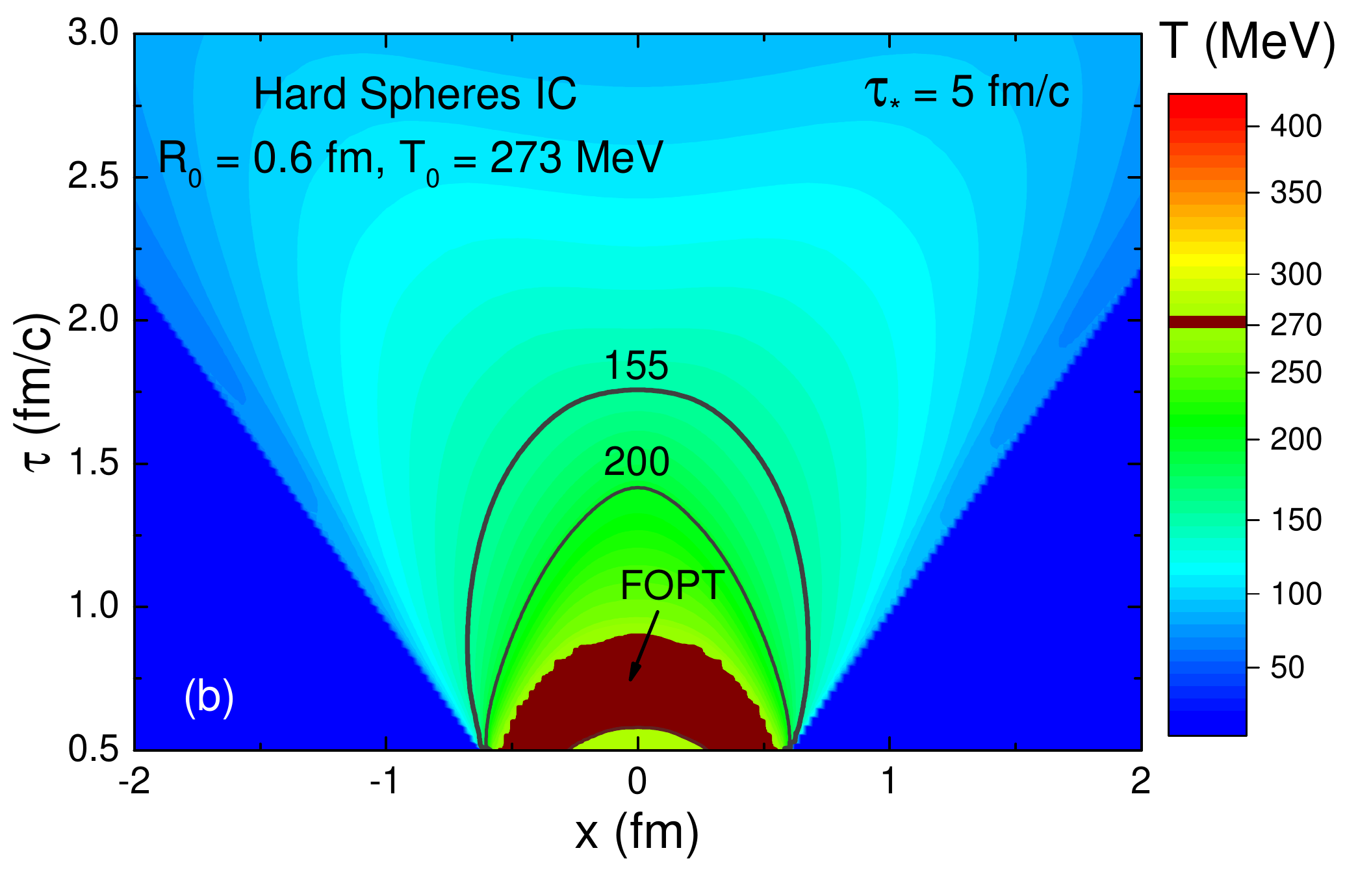}
\caption{The temperature profile of the central cell in the
  longitudinally boost invariant (2+1)-dimensional hydro evolution for
  $\text{p} \overline{\text{p}}$ collisions in the pure-glue initial
  state, the Yang-Mills scenario. A hard spheres overlap, transverse
  density profile with radius $R = 0.6 \,\fm$ is used as the initial
  condition. The normalization is fixed in order to yield an initial
  temperature of 273\,MeV in the central cell. (a) The $\tau$-dependence
  of the temperature is given for the central cell for different quark
  equilibration times: $\tau_* = 0$ (instant equilibration), $1\,\fm/c$
  (fast equilibration), $5 \,\fm/c$ (moderate equilibration),
  $10 \,\fm/c$ (slow equilibration), and for $\tau_* \to \infty$ (pure
  gluodynamic evolution).  (b) The temperature profile in the $x-\tau$
  plane for $\tau_* = 5 \,\fm/c$.}
\label{fig:pp-hydro}
\end{figure}

Figure~\ref{fig:pp-hydro} (a) shows the $\tau$-dependence of the
temperature in the central cell for different quark equilibration times:
$\tau_* = 0$ (instant equilibration), $1 \,\fm/c$ (fast equilibration),
$5\,\fm/c$ (moderate equilibration), $10 \,\fm/c$ (slow equilibration),
and $\tau_* \to \infty$ (pure gluodynamic evolution). In the pure
gluodynamic scenario, $\tau_* \to \infty$, the system spends a very long
time in the mixed-phase region. A fast quark equilibration shortens the
time period spent in the mixed phase significantly. Nevertheless, a
significant fraction of the system evolution takes place in the mixed
phase of the gluon-glueball deconfinement phase transition for a
moderately fast quark equilibration ($\tau_* = 5\,\fm/c$), as
illustrated by Fig.~\ref{fig:pp-hydro}b. Thus, significant effects of
the initial pure-glue state on electromagnetic and hadronic observables
are expected for this collision setup.

These results illuminate the future HESR-collider option with the
central PANDA experiment detector as an exciting upgrade for FAIR, a
promising option to search for even heavier glueballs and hadrons than
envisioned for the fixed target mode, and for other new exotic states of
matter. Of course, the results have been obtained within a largely
schematic calculation of $\overline{\text{p}} \text{p}$ collision
dynamics which employs ideal hydrodynamics and chemical equilibration of
quarks put in by hand.  Nevertheless, we hope that the possibility to
observe remnants of the first-order phase transition in pure
gluodynamics suggested by the result will motivate experimental
measurements at PANDA as well as more detailed theoretical studies based
e.g.\ on viscous hydrodynamics or partonic transport theory.

\subsection{Constituent antiquark-quark Drell- Yan processes and
  dilepton production}

Low-mass lepton pair production has raised the interest in the field for
decades. A quite robust theoretical understanding
\cite{Rapp:1999us,Rapp:2009yu,
  Endres:2015egk,Galatyuk:2015pkq,Staudenmaier:2017vtq,Linnyk:2015rco}
of dilepton production in heavy-ion collisions at various energies has
been gained. There dileptons play a special role as messengers from the
early stages, as penetrating probes.

\begin{figure}[t]
\hspace{-0.5cm}
\includegraphics[width=0.55\linewidth]{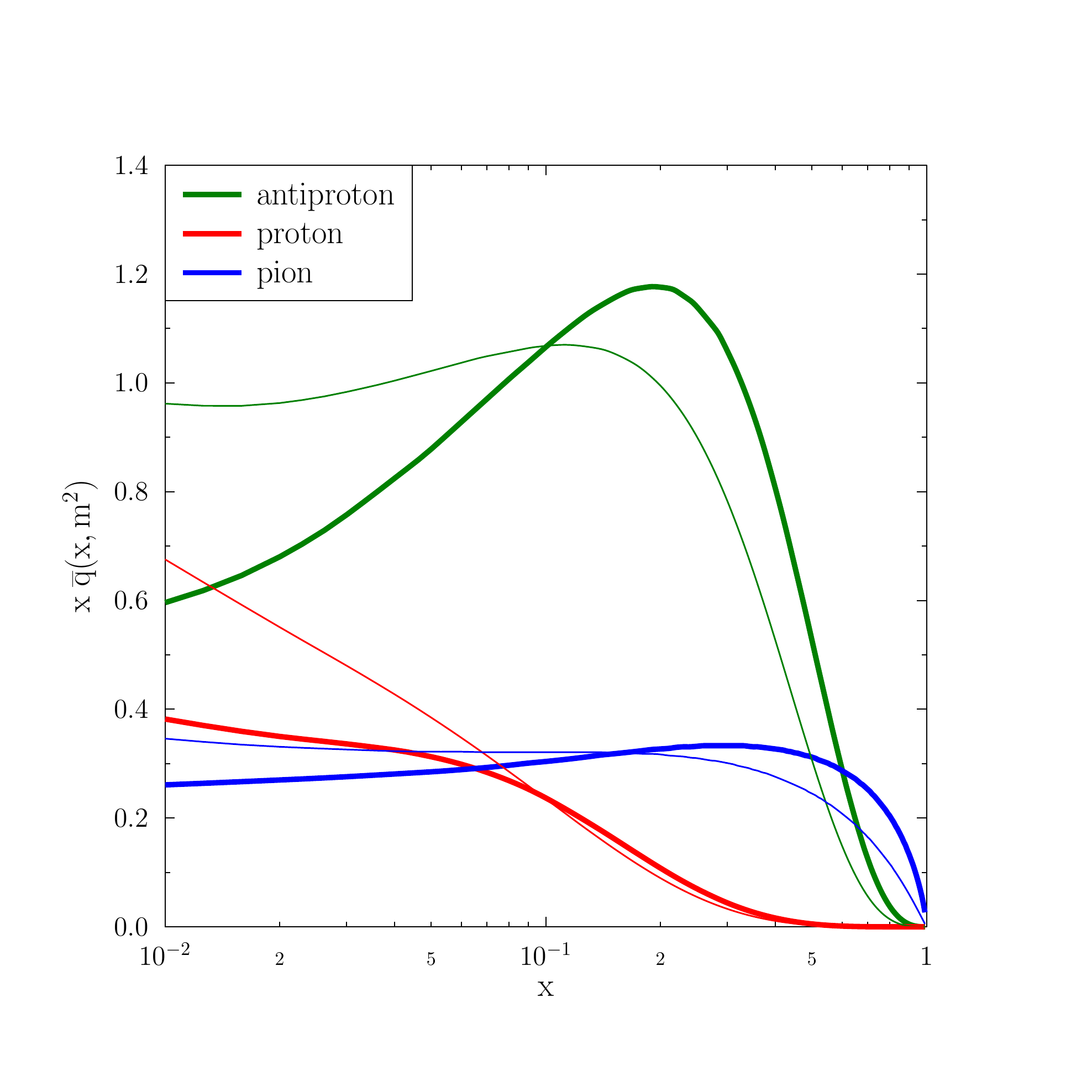} \hspace{-0.5cm}
\includegraphics[width=0.55\linewidth]{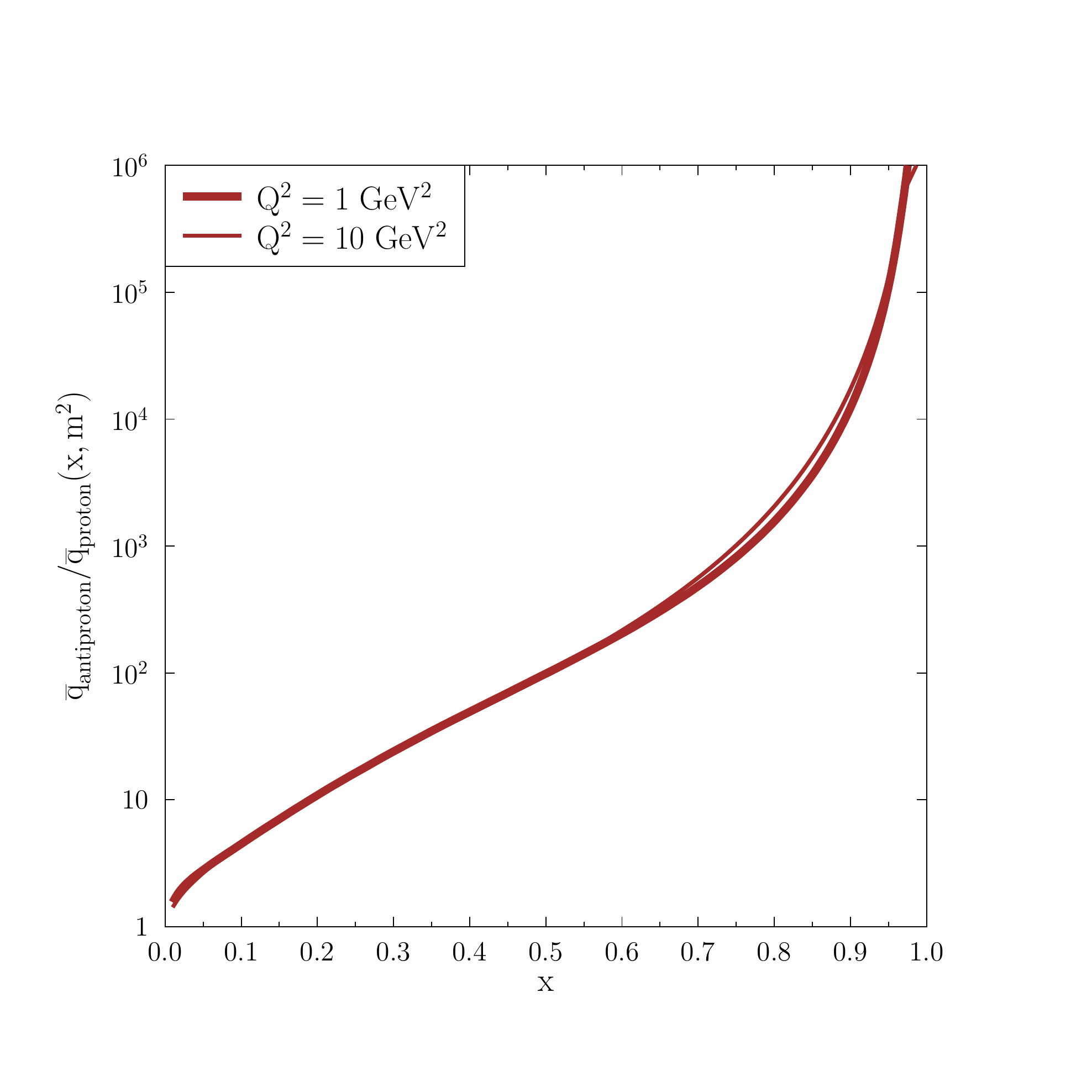}
\caption{Left: The GRV LO light-anti-quark parton-distribution functions
  (PDFs) at $m^2=10 \;\GeV^2$ (thin lines) and $m^2=1 \; \GeV^2$ (bold
  lines) for antiprotons, protons \cite{Dulat:2016}, and pions
  \cite{Gluck:1999}. Right: The ratio of the PDFs for light anti-quarks
  in antiprotons and in protons. For further details, see
  \cite{Spieles:1997ih,Spieles:1998wz,Spieles:2019dy}.}
\label{fig.pdfs}
\end{figure}

At large invariant dilepton masses the perturbative Drell-Yan (DY)
mechanism of QCD sets in. The minimal $M_{\text{DY}}^2$ value, where the
DY mechanism dominates, is not well known as it is difficult to separate
it experimentally from the contribution of charm and $\text{J}/\uppsi$
production. It seems that the DY pair production mechanism dominates at
$M_{\text{DY}}\geq 2\,\GeV$. For $\overline{\text{p}} \text{p}(A)$
collisions at intermediate energies, $M_{\text{DY}}$ should be lower
than for proton projectiles due to the lower background contributions to
lepton pairs from hadronic decay because of the presence of abundant
valence antiquarks as discussed above.

Higher-twist mechanisms may delay the inset of the leading-twist
contribution in the case of interactions with nuclei. Drell-Yan
production of dileptons at intermediate invariant masses by reactions
with secondary mesons and (anti-)baryons in p$A$ and $AA$ collisions has
been investigated in
\cite{Spieles:1997ih,Spieles:1998wz,Spieles:2019dy}. At the lower beam
energies of the beam-energy scan at RHIC, at FAIR, and at NICA, strong
contributions from secondary DY processes are predicted for low- and
intermediate-mass dilepton pairs due to the formation of mesons at time
scales $\lesssim 1 \,\fm/c$, i.e. during the interpenetration stage of
projectile and target. This implies that inflying primordial projectile
and target nucleons from the interpenetrating nuclei collide with just
newly formed mesons (e.g., $\rho$ and $\omega$ with constituent-quark
and constituent-antiquark masses of $\sim 300 \, \MeV$ each).

For $\overline{\text{p}}\text{p}$ and $\overline{\text{p}} A$ reactions,
the Drell-Yan production is enhanced already for the primordial
collisions due to the presence of \emph{valence} antiquarks of the
antiproton, as displayed in Fig.\ \ref{fig.pdfs} which shows the ratio
of light-antiquark PDFs in antiprotons and protons. With the HESR-C
$\overline{\text{p}}$p and $\overline{\text{p}}A$ collider discussed
here a direct assessment of the valence-quark valence-antiquark parton
distributions in $\overline{\text{p}} \text{p}$ and
$\overline{\text{p}} A$ collisions is accessible. The comparison of
dilepton production by proton and antiproton induced reactions provides
unique information on the relative role of initial- and final-state
mechanisms. In particular, by comparing dilepton production in proton
and antiproton fragmentation regions one may expect the maximal
difference between the two cases.

Calculations of the inclusive (integrated over transverse momenta) DY
production can be performed now in the NNLO DGLAP approximation. The
techniques were developed for calculations of transverse momentum
distributions which include effects of multiple gluon emissions (the
Sudakov form factor effects) and nonperturbative transverse momentum
distributions (TMD), see e.g. \cite{Angeles-Martinez:2015sea} and
references therein. TMDs contribute to the differential cross section
predominantly at small transverse momenta of the DY pairs and for
moderate masses of the pairs.

For large masses the theory works now very well providing a parameter
free description of $Z-$boson production at the LHC including the
transverse-momentum distribution.

\begin{figure}[t]
\includegraphics[keepaspectratio,height=4.5cm]{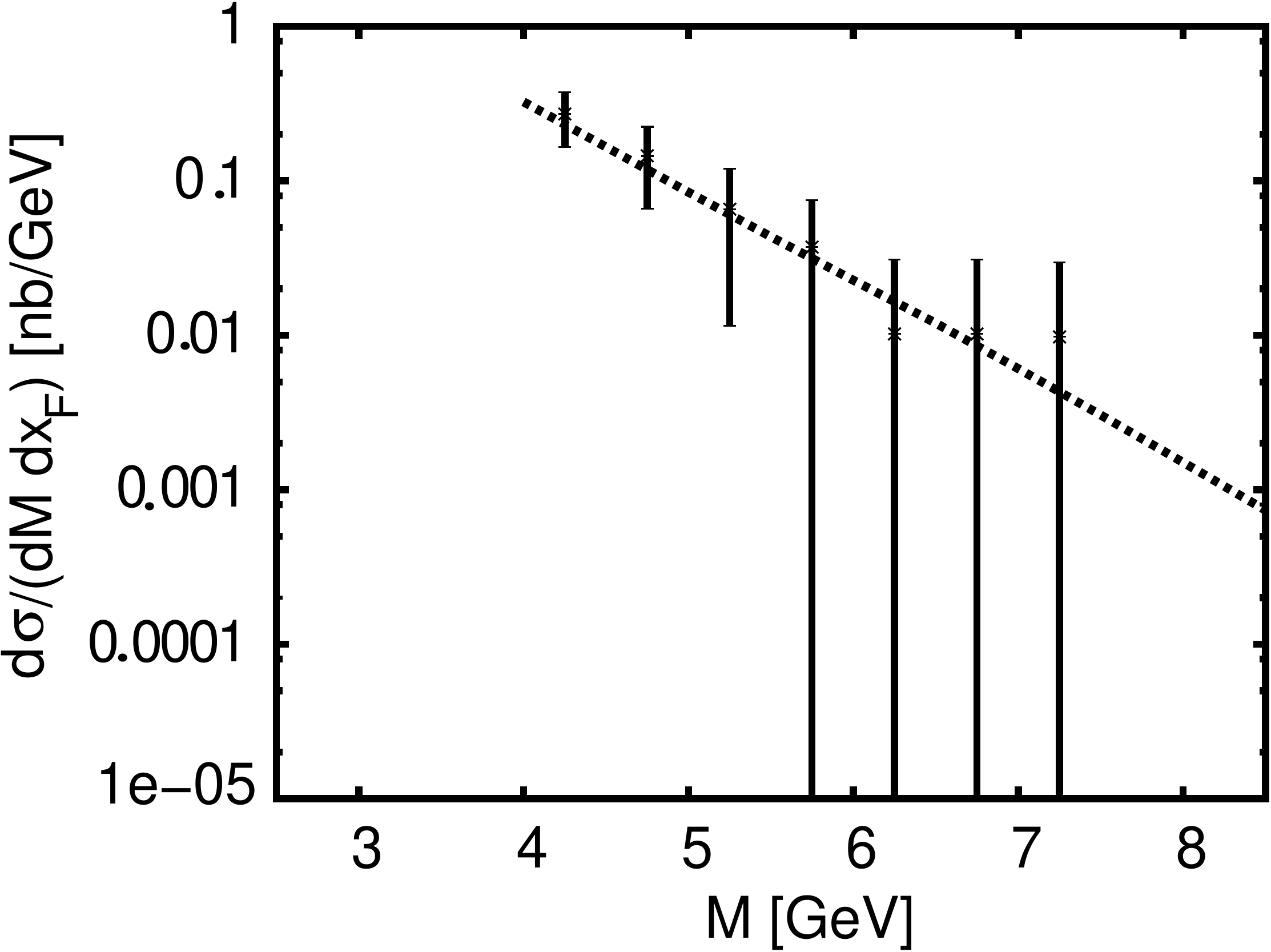} \hfill
  \includegraphics[keepaspectratio,height=4.5cm]{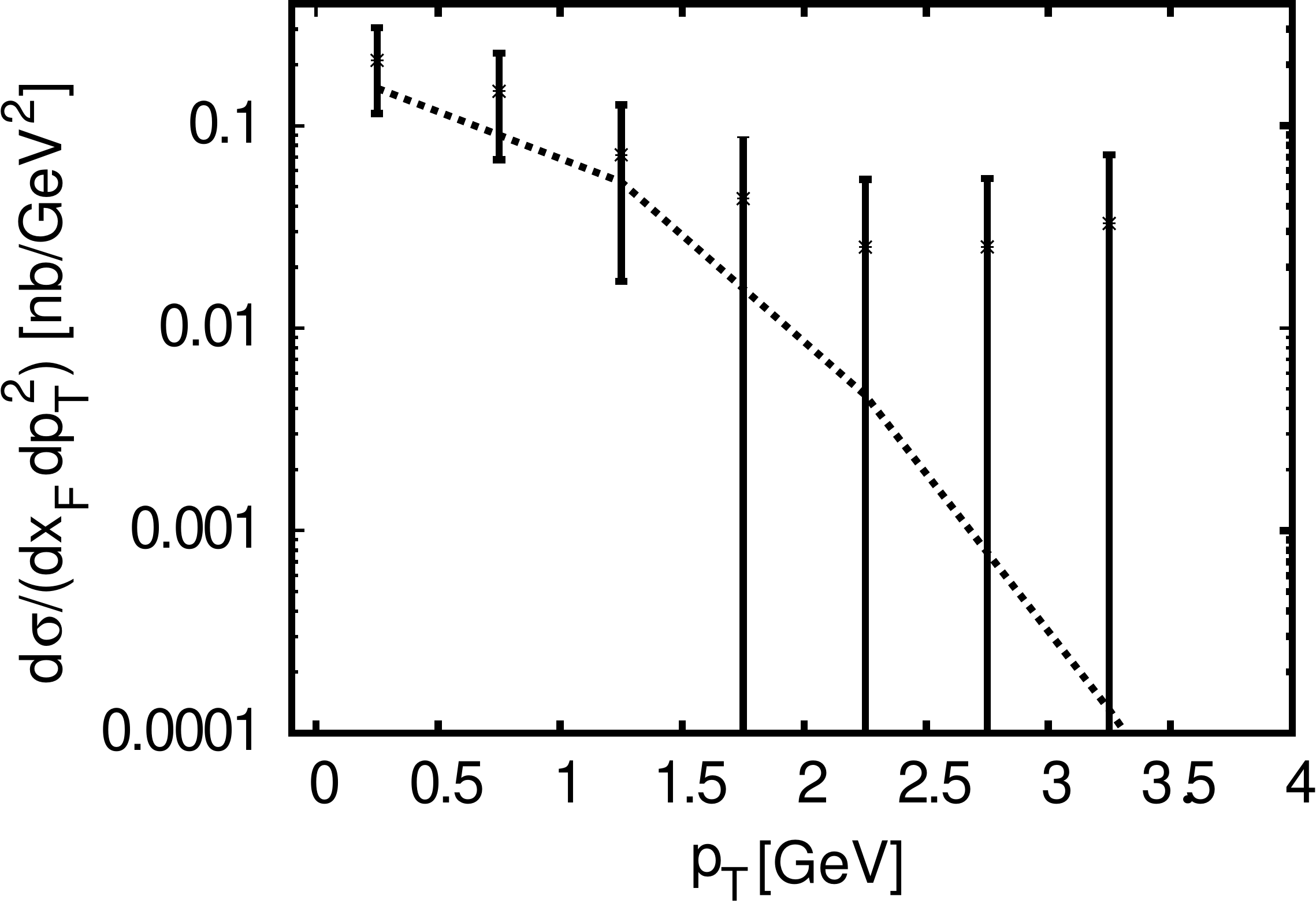}
  \caption{ Invariant-mass (left) and transverse-momentum spectra
    (right) of DY dimuon pairs in $\overline{\text{p}}\text{W}$ at
    $\sqrt{s}=15 \; \GeV$ collisions from the model described in
    \cite{Eichstaedt:2011ke} in comparison to data are from the E537
    collaboration \cite{Anassontzis:1987hk}. The used integrated PDF's
    are the MSTW2008LO68cl set \cite{Martin:2009iq}.}
  \label{fig.pbar-W-E537-PANDA}
\end{figure}

Corresponding high precision data for fixed target energies are very
limited especially for the antiproton projectiles. As an illustrative
example Fig.\ \ref{fig.pbar-W-E537-PANDA} shows the result of the model
for $\overline{\text{p}}\text{W}$ collisions of the E537 collaboration at
$\sqrt{s}=15 \; \GeV$ \cite{Anassontzis:1987hk}.

The data are compared with the model calculation
\cite{Eichstaedt:2011ke} which includes pQCD parton evolution, and the
TMD effects.

The model parameters have been fixed by using data on dimuon
transverse-momentum spectra in pp collisions at $\sqrt{s}=39 \, \GeV$
from the E866 collaboration in \cite{Webb:2003ps,Webb:2003bj}. This
model describes the dimuon production data without any adjustment of
model parameters quite well, in particular, the absolute cross section
in pd collisions from the E772 collaboration \cite{McGaughey:1994dx}, in
pCu collisions from the E605 collaboration \cite{Moreno:1990sf} at the
same collision energy, as well as in $\text{p}A$ collisions from the
E288 collaboration \cite{Ito:1980ev} and in pW collisions from the E439
collaboration \cite{Smith:1981gv} at $\sqrt{s}=27~\GeV$.

The presence of the abundant anti-quarks and the forward kinematics
experimental arm of PANDA at the HESR-C $\overline{\text{p}} \text{p}$
collider allows to determine the minimal $M^2$ for which the DY
mechanism works since the background contributions are expected to be
surpressed in comparison to pp collisions due to the above discussed
presence of the valence-anti-quarks in the $\overline{\text{p}}$. (Note
that the charm contribution is strongly suppressed at $x_p \ge 0.4$ as a
lepton in a charm decay carries, on average, only 1/3 of the D-meson
momentum (and even less for charm baryons, and that in addition PANDA is
designed to measure this contribution accurately).

The onset of factorization in the DY process with nuclei has still not
been explored, but will be very interesting. Indeed, as mentioned
already, the formation length in these processes is pretty small.
Hence, one could think of the process in a semiclassical way, for a
rather broad range of energies. In this picture, in the case of a
nuclear target, the antiproton may experience one or more (in)elastic
rescatterings on the target nucleons -- before it annihilates with a
proton into a dilepton pair. This effect can be taken into account
within transport models, e.g. GiBUU \cite{Buss:2011mx} calculations,
which are a good setup for the antiproton-nucleus dynamics. The
antiproton stopping shall lead to a softening of the invariant-mass and
transverse-momentum spectra of the dilepton pairs. It is also important
to include these effects in future calculations, as it influences the
conclusions on the in-medium modifications of the nucleon PDFs.

At high enough energies the formation time becomes large, but an
antiquark which is involved in the DY process can experience energy
losses growing quadratically with the path length~\cite{Baier:1996sk}.
Here one expects $p_T$ broadening $\propto A^{1/3}$. These studies, if
performed as a function of the atomic number and of the collision energy
at fixed $x_{\overline q}$, may allow to explore these important effects
in great detail.

\section{Final remarks}

The exciting science discussed here could be extended to much higher
energies and heavier states if -- in a later phase of FAIR -- one can
manage to reinject antiprotons from the HESR into one or two of the
higher energy main FAIR synchrotrons, SIS100 for
$\sqrt{s} \leq 60 \,\GeV$ and SIS300 for $\sqrt{s} \leq 180 \,\GeV$
$\overline{\text{p}} \text{p}$ collisions. Then
$\overline{\text{p}} \text{p}$ collisions can be synchronized to run
effectively and with high luminosity in a collider mode too. Also,
crossing beams of SIS100 with SIS300, e.g. $45 \,A \GeV$ heavy ions in
the SIS300 colliding with $30 \,\GeV$ antiprotons (or protons), or with
ions of $15\, A\GeV$ in the SIS100 can be envisioned. Also other
asymmetric collisions maybe feasible, e.g., $90\,\GeV$
antiprotons/protons (SIS300) colliding with $15 \, A\GeV$ heavy ions (or
with $30\,\GeV$ protons or antiprotons), in the SIS100, when a common
interaction region between SIS100 and SIS300 can be established.

\section*{Acknowledgments}

We thank M.~Cacciari and R.~Vogt for discussions on charm and beauty
production in $\text{p} \overline{\text{p}}$ scattering, A.\ Motornenko
for pointing at recent work on correlations as well as L.\ Schmitt and
U.\ Wiedner for providing detailed information about the present status
of the PANDA detector. The research by L.F.\ and M.S.\ was supported by
the US Department of Energy Office of Science, Office of Nuclear Physics
under Award No. DE-FG02-93ER40771.  M.S.\ also acknowledges support by
the A.\ von Humboldt foundation.  A.L.\ acknowledges partial financial
support by the Helmholtz International Center (HIC) for FAIR.  H.v.H.\
acknowledges the support from Frankfurt Institute for Advanced
Studies~(FIAS). H.St.\ acknowledges the support through the Judah
M. Eisenberg Laureatus Chair by Goethe University and the Walter Greiner
Gesellschaft, Frankfurt. A.\ Larionov acknowledges support by the German
Federal Ministry of Education and Research (BMBF), Grant No. 05P18RGFCA.

\section*{Appendix: PANDA detector}
\label{sect:app-panda-dect}

The PANDA detector at FAIR can be used without modifications as a
midrapidity detector in a future antiproton-proton/ion HESR-collider:
primary high energy proton beams from the SIS18 at GSI are injected
clockwise or counterclockwise into the HESR-C, see the dark-blue lines
in Fig.\ \ref{fair}, and collide with the $\overline{\text{p}}$'s
created at the SIS100 synchrotron complex, collected in the collector
ring and from there filled counterclockwise or clockwise into the High
Energy Storage Ring (HESR); $\overline{\text{p}}$ and p then can be
simultaneously accelerated in the magnetic fields in the HESR. The
$\overline{\text{p}}$'s shall collide with the proton beam injected from
SIS18 in the middle of the PANDA Detector. The PANDA collaboration
focuses, in the presently planned fixed target mode with
$\sqrt{s_{\text{max}}} \sim 5.5\, \GeV$, on hadron spectroscopy, search
for exotic hadrons, hadrons in media, nucleon structure and
``(multi-)hyper-, strangelets, charmed and future beauty exotic
nuclei''. To fully exploit PANDA and the HESR, we present the next
generation physics interest for the high energy antiproton-proton
collider physics, which can be realized in a highly cost effective
manner by just adding one simple beam line from the existing SIS18
synchrotron to the HESR, and by using the versatile, basically
unmodified PANDA detector, to provide precise trajectory reconstruction,
energy and momentum measurements and an efficient particle
identification system, for a momentum range from 1.5 to $15\,\GeV/c$,
with peak luminosities up to $\sim 10^{31} \text{cm}^{-2}\text{s}^{-1}$,
and a number of stored antiprotons and protons, $N\sim 10^{11}$.  PANDA
provides nearly full coverage of the solid angle for the produced
particles, together with good particle identification and high momentum
($\Delta p/p<1\%$) and angular resolutions for both, charged particles
and photons.

PANDA is subdivided into the target spectrometer (TS), consisting of a
large solenoid around the interaction region, and a forward spectrometer
(FS), based on a dipole, to momentum-analyze the forward-going particles,
allowing for the full angular coverage, for a wide range of
energies.

The tracking systems are, from the inside outwards: the
Micro-Vertex-Detector (MVD) in the TS yields a precise determination of
secondary decay vertices of short-lived particles, the Straw Tube
Tracker (STT) serves as central tracking device in the magnetic field
providing a good momentum resolution.

A large area planar Gaseous Electron Multiplier GEM-Tracker provides
trajectory position resolution of $< 100 \, \upmu \text{m}$ for the
forward going tracks in the TS.

The Forward Tracker (FT) is based on the same type of straws as the STT
and is designed for momentum analysis of charged particles deflected in
the field of the PANDA dipole magnet aiming at a momentum resolution
better than 1\%.

Electromagnetic Calorimetry of PANDA allows for high count rates - the
target spectrometers consist of $>15000 \,\text{PbWO}_4$ crystals for an
excellent pion/electron discrimination for momenta above $0.5 \, \GeV/c$
and is designed to detect photons in the energy range between
$10 \, \MeV$ and $15 \,\GeV$ with a resolution $\sim 2\%$.

The identification of charged particles with extreme accuracy is one of
the key requirements for unveiling many aspects of both, the fixed
target program as well as the new collider physics program envisioned
here for PANDA in the decade 2030-2040.

The PANDA Detector features dedicated particle identification (PID)
systems which classify particle species over a wide kinematic range, in
addition to $\dd E/\dd x$ measurements from tracking and information
from electromagnetic calorimetry, the Detection of Internally Reflected
Cherenkov light (DIRC) in the TS, a Time Of Flight System, a Muon Detection
System and a Ring Imaging Cherenkov Detector in the FS.

This fully equipped PANDA Detector System, as now designed for the fixed
target mode of the FAIR-MSV, is also a formidable multi-purpose
instrument when used as a midrapidity HESR-collider detector:

The unchanged fixed target design's forward arm can be used for both the
antiproton side - but also for the proton side in the collider mode, if
the antiprotons' injection direction into HESR can be switched from
upwards to downwards - and, simultaneously for the protons, from
downwards to upwards.

This can be accomplished in a cost effective manner, i.e. without
changing the location of the forward arm and without adding a second
spectrometer, in the following manner: both injection lines into the
HESR-collider, namely
\begin{itemize}
\item the new short proton beam line, shown in red in Fig. 1, which
  comes directly from the SIS18 transfer beamline to SIS100 and
\item the antiproton beam line, coming from the SIS100-antiproton
  production target, through the CR,
\end{itemize}
will be equipped not with only one, but with two deflector switches
each, respectively, into the HESR-Collider, allowing for clockwise and
counterclockwise rotation of both beams, protons and antiprotons,
similar as the LHCb detector at CERN.

This enables the full HESR-C physics program - with minor adjustments to
HESR - as described in the present paper - with an unmodified PANDA
detector - a most cost-effective proposal for realizing most exciting
scientific goals.

\begin{flushleft}
\bibliographystyle{num-hvh-tit}
\bibliography{pandacollit}
\end{flushleft}

\end{document}